\documentclass[aoas]{imsart}

\RequirePackage{amsthm,amsmath,amsfonts,amssymb}
\RequirePackage[authoryear]{natbib}
\usepackage[dvipsnames]{xcolor}
\usepackage{physics}
\usepackage{hyperref}
\usepackage{aasmacros}
\usepackage{tikz}
\usepackage{algorithm}
\usepackage{algpseudocode}
\usepackage{mathdots}
\usepackage{yhmath}
\usepackage{cancel}
\usepackage{color}
\usepackage{array}
\usepackage{multirow}
\usepackage{amssymb}
\usepackage[dvipsnames]{xcolor}
\usepackage{gensymb}
\usepackage{tabularx}
\usepackage{fontspec}
\usepackage{extarrows}
\usepackage{booktabs}
\usetikzlibrary{fadings}
\usetikzlibrary{patterns}
\usetikzlibrary{shadows.blur}
\usetikzlibrary{shapes}
\usepackage{comment}

\RequirePackage{graphicx}

\startlocaldefs
\theoremstyle{plain}

\newcommand{\rev}[1]{#1}

\newcommand{\dmunits}{pc\,cm$^{-3}$}

\newcommand{\pcc}{$P_{\text{C}}$}
\newcommand{\pccl}{probability of coincidence}
\DeclareMathOperator{\sinc}{sinc}
\endlocaldefs

\begin{document}

\begin{frontmatter}
\title{K-Contact Distance for Noisy Nonhomogeneous Spatial Point Data with application to Repeating Fast Radio Burst sources}
\runtitle{K-Contact Distance for Noisy NHPP}
\begin{aug}
\author[MU]{\fnms{A. M.}~\snm{Cook}\ead[label=ac]{amanda.cook@mail.mcgill.ca}\orcid{0000-0001-6422-8125}},
\author[DoSS]{\fnms{Dayi}~\snm{Li}\ead[label=dl]{dayi.li@mail.utoronto.ca}\orcid{0000-0002-5478-3966
}},
\author[DoSS]{\fnms{Gwendolyn M.}~\snm{Eadie}\ead[label=ge]{gwen.eadie@utoronto.ca}\orcid{0000-0003-3734-8177}},
\author[SFU]{\fnms{David C.}~\snm{Stenning}\ead[label=ds]{david\_stenning@sfu.ca}\orcid{0000-0002-9761-4353}},
\author[YU]{\fnms{Paul}~\snm{Scholz}\ead[label=ps]{pscholz@yorku.ca}\orcid{0000-0002-7374-7119}},
\author[SFU]{\fnms{Derek}~\snm{Bingham}\ead[label=db]{dbingham@sfu.ca}},
\author[DoSS]{\fnms{Radu}~\snm{Craiu}\ead[label=rc]{radu.craiu@utoronto.ca}\orcid{0000-0002-1348-8063}},
\author[UCSC]{\fnms{B. M.}~\snm{Gaensler}\ead[label=bg]{gaensler@ucsc.edu}\orcid{0000-0002-3382-9558}},
\author[MITK]{\fnms{Kiyoshi W.}~\snm{Masui}\ead[label=km]{kmasui@mit.edu}\orcid{0000-0002-4279-6946}},
\author[UVA]{\fnms{Ziggy}~\snm{Pleunis}\ead[label=zp]{z.pleunis@uva.nl}\orcid{0000-0002-4795-697X}},
\author[DAA]{\fnms{Antonio}~\snm{Herrera-Martin}\ead[label=ah]{antonio.herreramartin@utoronto.ca}\orcid{0000-0002-3654-4662}},
\author[MU]{\fnms{Ronniy C.}~\snm{Joseph}\ead[label=rcj]{ronniy.joseph@mcgill.ca}\orcid{0000-0003-3457-4670}}
\author[DAA]{\fnms{Ayush}~\snm{Pandhi}\ead[label=ap]{ayush.pandhi@mail.utoronto.ca}\orcid{0000-0002-8897-1973}},
\author[MU]{\fnms{Aaron B.}~\snm{Pearlman}\ead[label=abp]{aaron.b.pearlman@physics.mcgill.ca}\orcid{0000-0002-8912-0732}},
\and
\author[UCSC]{\fnms{J. Xavier}~\snm{Prochaska}\ead[label=x]{xavier@ucolick.org}\orcid{0000-0002-7738-6875}}
\address[MU]{Department of Physics, McGill University\printead[]{ac,rcj,abp}}
\address[DoSS]{Department of Statistical Science, University of Toronto\printead[presep={,\ }]{dl,ge,rc}}
\address[DAA]{David A. Dunlap Department of Astronomy \& Astrophysics, University of Toronto\printead[presep={,\ }]{ac,ah,ap}}
\address[SFU]{Department of Statistics and Actuarial Science, Simon Fraser University\printead[presep={,\ }]{ds,db}}
\address[YU]{Department of Physics and Astronomy, York University
\printead[presep={,\ }]{ps}}
\address[UCSC]{Department of Astronomy and Astrophysics, University of California, Santa Cruz\printead[presep={,\ }]{bg,x}}
\address[MITK]{MIT Kavli Institute for Astrophysics and Space Research, Massachusetts Institute of Technology\printead[presep={,\ }]{km}}
\address[UVA]{Anton Pannekoek Institute for Astronomy, University of Amsterdam\printead[presep={,\ }]{zp}}
\end{aug}

\begin{abstract}
This paper introduces an approach to analyze nonhomogeneous Poisson processes (NHPP) observed with noise, focusing on previously unstudied second-order characteristics of the noisy process. Utilizing a hierarchical Bayesian model with noisy data, we estimate hyperparameters governing a physically motivated NHPP intensity. Simulation studies demonstrate the reliability of this methodology in accurately estimating hyperparameters. Leveraging the posterior distribution, we then infer the probability of detecting a certain number of events within a given radius, the $k$-contact distance. We demonstrate our methodology with an application to observations of fast radio bursts (FRBs) detected by the Canadian Hydrogen Intensity Mapping Experiment's FRB Project (CHIME/FRB). This approach allows us to identify repeating FRB sources by bounding or directly simulating the probability of observing $k$ physically independent sources within some radius in the detection domain, or the \textit{\pccl} (\pcc). The new methodology improves the repeater detection \pcc\, in \rev{91}\% of cases when applied to the largest sample of previously classified observations, with a median improvement factor (existing metric over \pcc\, from our methodology) of $\sim$ \rev{4800}. 
\end{abstract}


\end{frontmatter}

\section{Introduction}
\subsection{Nonhomogeneous Poisson Point Processes}
To learn about the nature of unknown astrophysical phenomena, astrophysicists can search for connections between different sources or across populations. 
A question of interest to astrophysicists is often: How do we estimate the likelihood of two or more spatially coincident, but separated in time, events being physically related when measurement uncertainty on event positions is non-negligible, the probability of detection varies on the sky, and the data is noisy? These connections, in particular spatial coincidences, can be ambiguous owing to instrumentation limitations, observational biases, and the sheer number of objects in the Universe. Such complicating factors often lead to non-homogeneous spatial point processes that are observed with noise. 


Considerable effort has gone into modelling noisy nonhomogeneous Poisson processes \citep[NHPPs;][]{van2000markov}. 
For example, \cite{477bb175-56f0-3455-81ae-9da35286a358} presented an edge-corrected deconvoluting kernel estimator for the intensity of noisy and nonhomogeneous spatial data and a bandwidth selection procedure for that kernel estimator.  \cite{10.1214/10-ba504} also estimated the two dimensional intensity of an NHPP observed with Gaussian noise. In their work, the authors derived the likelihood function for the intensity of an NHPP observed with noise, also considering the effect of events which, due to the stochastic error, are moved inside or outside of the observed domain and hence cause miss-estimation of the intensity near these boundaries. 
While these approaches allow for descriptions of an NHPP intensity function from noisy data, neither methods are used to make predictive inferences about second order characteristics. 

\rev{Alternatively}, \cite{lund1999} and \cite{Bar-Hen_Chadoeuf_Dessard_Monestiez_2013} made inferences based on observations from noisy NHPPs with numerically estimated second order characteristics (e.g., the nearest-neighbor function). In those works, however, the authors were interested in recovering these second order characteristics for the underlying unobserved NHPP from the noisy observations, and not those of the observed noisy process itself. For astrophysicists that are interested in quantifying the probability of observing spatially coincident and separate in time, but ultimately physically unrelated noisy events, the unobserved NHPP is not the process of interest. 


In this paper, we propose a hierarchical Bayesian model for noisy observations to estimate an NHPP intensity function with the purpose of making predictive inference on the \textit{noisy process}. Existing statistical methodologies seek only to make inferences on the underlying or true process given noisy observations of it. We first construct a hierarchical Bayesian model of a parameterized Poisson intensity function that is motivated by the data-collecting instrument  (in our application, a telescope). We then use this intensity function to estimate the probability of observing two or more ($k \geq 2, k \in \mathbb{N}$) events 
within a given observed radius in the noisy NHPP assuming they are physically independent: the $k$-contact distance. This probability is estimated both using direct simulation and a novel and analytic probability bound we derive on the $k$-contact distance for the noisy NHPP. 
This then allows us to estimate the probability that the $k$ physically independent events are within some radius (in the observation domain) of one another and, if sufficiently improbable, give evidence that two or more events may plausibly be caused by the same source, meaning the same astrophysical body or system, as opposed to sources that sit nearby each other in the observation domain. 
While this methodology is geared towards astrophysical observations with significant localization uncertainty, the common application of NHPPs to fields like ecology and epidemiology suggests that the method may be beneficial for research in other disciplines, for example, hotspot detection.
\subsection{Motivating Application}
Fast radio bursts (FRBs) are short, bright, extragalactic radio signals of unknown origin. Our motivation for the proposed methodology is to distinguish clusters of physically unrelated but spatially coincident FRB sources from true \textit{repeating} FRB sources, which we describe next.

Many source models have been proposed for FRBs, often drawing connections to exotic populations like neutron stars and black holes \citep[e.g.,][]{2019PhR...821....1P, 2022A&ARv..30....2P, 2022arXiv221203972Z}. While the majority of FRBs have been observed to happen once, a minority are observed to repeat. That is, multiple bursts (signals) from one location in the sky are observed, presumably from the same source \citep[e.g.][]{2016Natur.531..202S, 2019Natur.566..235C, 2019ApJ...885L..24C, 2023ApJ...947...83C, 2020ApJ...891L...6F, 2022Natur.606..873N}. 
The discovery of repeating sources, or \textit{repeaters}, suggests that at least some of the extremely luminous bursts need to be sourced from non-cataclysmic events where the source is not consumed as a result of the production of the burst. An example of a cataclysmic source would be, e.g., a neutron star collapsing into a black hole \citep{2014A&A...562A.137F} or mergers of compact objects, e.g., two neutron stars \citep{2013PASJ...65L..12T}. Non-cataclysmic source theories can involve, e.g., interactions of compact objects while they periodically orbit eachother \citep{2021ApJ...917...13S}.  
There are still many proposed theories for this repetition, but the non-cataclysmic nature of the events significantly narrows the number of possible source models for these repeaters. 

Astronomers have detected and published 790 distinct sources of FRBs on the sky, of which 56 are known repeaters\footnote{\url{https://www.wis-tns.org/} accessed Sept 25th, 2024}.    
The most prolific repeaters have been observed to repeat hundreds to thousands of times, while dozens have just two or three total observed bursts \citep[e.g. ][]{2021Natur.598..267L, 2023ApJ...947...83C}. 
While there is some evidence that repeating FRBs are a distinct population from apparent non-repeating FRBs \citep{2021ApJ...923....1P}, it is not known definitively whether there are two separate populations or if all FRBs repeat \citep{2023arXiv230617403J,2023arXiv230615505K}.

Most FRB sources detected to date have been discovered by 
the Canadian Hydrogen Intensity Mapping Experiment (CHIME). 
 A traditional telescope will point to a small patch of the sky and track that position over time. CHIME is a transit radio telescope, meaning it has no moving parts and simply observes the sky passing overhead \citep{2022ApJS..261...29C,overview}. 

The stationary design of CHIME has observational benefits, like a large field-of-view and daily monitoring of sources, as well as challenges, such as exposure that is uneven on the sky. 
At the same time, CHIME has limited angular resolution, that is, ability to pinpoint exactly where the sources are located on the sky. The FRB project of CHIME (CHIME/FRB) observes FRBs with 99.7\% localization (sky position) uncertainty regions that span, at best, a few tens of square arcminutes\footnote{For a sense of scale, the full Moon subtends $\sim$30 arcminutes, and one arcminute is, by definition, \rev{1/60 of a degree.}}, or at worst a few square degrees. 
FRBs are found within galaxies, and it is physically reasonable that multiple independent FRB sources could be found in a localization region of this size since there are typically many galaxies (which could host FRBs) within these regions
\footnote{The well-known Hubble Ultra Deep Field image, in which every non-spiky point of light is a galaxy ($\sim 10,000$ such galaxies), is of comparable size to CHIME/FRB's smallest (best constrained) localization regions \citep{2006AJ....132.1729B}.}. 
In summary, a CHIME/FRB detection does not provide the precise position of a source, and as CHIME/FRB continues to detect more FRBs, the probability that two or more spatially independent FRB sources have observed concordant positions becomes non-negligible. This remains true even when considering the third \rev{`}resolving' dimension of FRBs, dispersion measure, which we introduce in Section~\ref{sec:coordinates}. Ongoing upgrades to the experiment will increase the localization precision for FRBs detected by CHIME/FRB \citep{2024arXiv240207898L}.  This increased localization precision usually allows for probabilistic selection between potential host galaxies, using, e.g., the `\textit{Probabilistic Association of Transients to their Hosts}' routine \citep[PATH; ][]{2021ApJ...911...95A}. However, there are other poorly-localized transient events whose potential association with FRBs would inform source models, such as gravitational wave sources and gamma-ray bursts \citep[e.g.,][]{2023NatAs...7..579M, 2024arXiv240409242C}. 

Application of the methodology that we propose will allow observers to make a probabilistic statement about the certainty of physical association between clusters of $k$ bursts observed with noise, i.e. potential repeaters, distinguishing them from independent sources that share overlapping localization regions on the sky. If CHIME/FRB surveyed all points in the Universe equally, making the probability of detecting FRBs spatially uniform, this process could be modeled as a three dimensional, homogeneous Poisson point process. Instead, there is uneven exposure and sensitivity in the intensity of FRB detections which is a non-trivial combination of spherical geometry, the luminosity distribution of FRBs, and instrumental effects. 

The net effect is an NHPP FRB detection intensity that must be empirically estimated from noisy telescope detections. NHPP processes are not only applicable to data from CHIME/FRB; NHPPs were first applied to FRB detections by \cite{2017AJ....154..117L}, who model the rate of FRBs.  

This article is organized as follows: in Section \ref{sec:model}, we specify a hierarchical model for observed noisy event locations from an NHPP. 
In Section \ref{sec:inference}, we outline the inference on the probability of observing $k$ events within a given volume in the observation domain. In Section \ref{sec:intensity}, we apply this model to data from CHIME/FRB, and derive a parametrization for CHIME/FRB's intensity function on the sky. In Section \ref{sec:performance}, we describe the implementation and performance of a method based on Markov chain Monte Carlo (MCMC) sampling for estimating the hyperparameters' posterior distribution. In Section \ref{sec:results}, we implement the proposed methodology on our application. We also estimate, via direct simulation and an analytic upper bound, the \pccl\,(\pcc) for each of CHIME/FRB's published repeaters and compare the probabilities to 
the existing measure of cluster significance used by \cite{2023ApJ...947...83C}. 
\section{Hierarchical Model for NHPP Observed with Noise}
\label{sec:model}

\begin{table}
\centering
\begin{tabular}{cl}
\hline
Symbol & Description \\ \hline
$\mathbf{Y}$ & point process of observed (noisy) positions \\
$\boldsymbol{y}_1, \ldots,  \boldsymbol{y}_n$ &  Realization of observed positions, from random process $\mathbf{Y}$  \\
$\boldsymbol{\varepsilon}_i$ & spatial perturbation vector due to measurement error\\
$n$ & Number of observed events \\
$\mathbf{X}$ & Nonhomogeneous poisson process (true positions)  \\
 $\boldsymbol{x}_1, \ldots, \boldsymbol{x}_n$ & Realization of true positions, from random process $\mathbf{X}$ \\
   $\Lambda$& Intensity function of NHPP $\mathbf{X}$\\
   $\boldsymbol{\theta}$ & Hyperparameter vector governing $\Lambda$\\
 $D$ & Domain on which positions are defined\\
 $k$ & Number of events in cluster\\
 $r$ & Radius of the minimal bounding sphere the cluster of events \\
 $s$ & General event location in the domain\\
  $s_0$ & Center of minimal bounding sphere cluster of events \\
 \hline
\end{tabular}
\caption{\label{tab:not} Definitions of application-independent variables used throughout this text.}
\end{table}

Let $\mathbf{X} \subseteq D \subseteq \mathbb{R}^d$ be an NHPP (we are concerned specifically with $d=3$ for our application given the event positions are observed in three dimensions, but remain general for now) with intensity function $\Lambda(\mathbf{s}; \boldsymbol{\theta}): D \rightarrow [0,\infty)$, that is, 
\begin{align}
    \mathbf{X} \sim  \text{NHPP}(\Lambda(\mathbf{s}; \boldsymbol{\theta})),
\end{align}
where $\Lambda$ is parameterized by vector $\boldsymbol{\theta}$, and $\Lambda(\mathbf{s}; \boldsymbol{\theta})$ is defined for any event location $\mathbf{s}$ in the bounded domain $D$ of $\mathbf{X}$. 
By the definition of a Poisson process and its intensity function, we must have that (i) $\Lambda(\mathbf{s}, \boldsymbol{\theta})$ is a Borel measure, (ii) given any bounded Borel set $B\subseteq D$ the probability mass function for number of observed events in $B$, $N(B)$, is Poisson distributed with mean $\Lambda(B; \boldsymbol{\theta})$, and (iii) for any $j$ disjoint bounded Borel sets $B_1,\ldots, B_j$, the random variables $N(B_1), \ldots, N(B_j)$ are independent \citep{van2000markov}. 
Following \cite{10.1214/10-ba504}, but under the simplifying assumption that events can not jump into or out of the domain due to stochastic error in our model, we also assume that no two events can share the same position, i.e., $\mathbf{X}$ is a \textit{simple} point process or $\Lambda$ is atomless. This is a null assumption in our application.
Given these properties, conditional on the number of events $n$ in $D$, the  locations $\boldsymbol{x}_1, \ldots, \boldsymbol{x}_n$ are independent draws from $\lambda(\boldsymbol{s}| \boldsymbol{\theta}) = \frac{1}{C} \Lambda(\boldsymbol{s}; \boldsymbol{\theta})$, where $C\in\mathbb{R}$ is the constant which normalizes $\Lambda(\boldsymbol{x}_i, \boldsymbol{\theta})$ to a density over $D$ \citep{KOTTAS20073151,10.1214/10-ba504}.
That is, the joint distribution of $n$ and $\mathbf{X} = \{\boldsymbol{x}_1, \ldots, \boldsymbol{x}_n\}$, conditional on $\boldsymbol{\theta}$ is:
\begin{align}
    L( \boldsymbol{\theta}| n; \boldsymbol{x}_1, \ldots, \boldsymbol{x}_n) \propto \exp(-\int_D \Lambda(\mathbf{s}; \boldsymbol{\theta}) \dd \mathbf{s}) \frac{\prod_{i=1}^n \Lambda(\boldsymbol{x}_i; \boldsymbol{\theta})}{n!}. \label{eqn:likst2}
\end{align}
The above follows from the conditional likelihood of $\boldsymbol{x}_1, \ldots, \boldsymbol{x}_n \propto \prod_i\lambda(\boldsymbol{x}_i| \boldsymbol{\theta})$ and probability mass function for the number of observed events, $n$, being
\begin{align}
    f(n|\Lambda, \boldsymbol{\theta}) =  \frac{\left(\int_D \Lambda(\boldsymbol{s}; \boldsymbol{\theta}) \dd \boldsymbol{s}\right)^n}{n!}\exp(-\int_D \Lambda(\mathbf{s}; \boldsymbol{\theta}) \dd \mathbf{s}). 
\end{align}
We derive the specific form of $\Lambda$ for the CHIME/FRB experiment in Section \ref{sec:intensity}, but this is left general at this stage.

Let $\mathbf{Y} = \{\boldsymbol{y}_1,\dots,\boldsymbol{y}_n\}$ be a point process of $n$ noisy event locations, which are a perturbed set of the $n$ true event locations from $\mathbf{X}$, $\boldsymbol{y}_i = \boldsymbol{x}_i+ \boldsymbol{\varepsilon}_i$, where $\boldsymbol{\varepsilon}_i$ is a random vector in $\mathbb{R}^d$ that describes the spatial perturbation due to measurement error. We do not assume this is Gaussian, as in previous works, but instead leave the error distribution general. 
The distribution of $\boldsymbol{\varepsilon_i}$ for our application is empirically estimated, and the details are deferred to Section~\ref{sec:noise}. 
 Our hierarchical model can thus be summarized as follows:
\begin{align*}
  \text{Stage I} \hspace{4mm} & \boldsymbol{y}_i | \boldsymbol{x}_i,\boldsymbol{\varepsilon_i} \mathop{\sim}^{\text{ind}} f(\cdot) \\
\text{Stage II} \hspace{4mm} &  \boldsymbol{x}_i|n \mathop{\sim}^{\text{ind}}  \lambda( \mathbf{s}, \boldsymbol{\theta})  \\
  \text{Stage III} \hspace{4mm} & n| \Lambda \sim \text{Pois}\left(\int_D \Lambda(\mathbf{s}, \boldsymbol{\theta}) \dd \mathbf{s}\right).
\end{align*}
A summary of the notation we use is given in Table \ref{tab:not}. We can thus write the full likelihood function of the above model as: 
\begin{align} 
    L(n; \boldsymbol{x}_1, \ldots, \boldsymbol{x}_n | \boldsymbol{\theta}) =  \frac{1}{n!}\exp(-\int_D \Lambda(\mathbf{s}, \boldsymbol{\theta}) \dd \mathbf{s})\left[ \prod_{i=1}^n \Lambda(\boldsymbol{x}_i, \boldsymbol{\theta}) \right] \left[ \prod_{i=1}^{n} f(\boldsymbol{y}_i| \boldsymbol{x}_i) \right], \label{eqn:likelihood}
\end{align}
where $f(\boldsymbol{y_i}| \boldsymbol{x_i})$ is the p.d.f. in Stage I evaluated at $\boldsymbol{y_i}$. 

\section{Bounding the Probability of Coincidence for Noisy NHPPs}

\label{sec:inference}
We want to quantify the likelihood of observing $k$ noisy event locations close together in our observation domain assuming the model outlined in Section \ref{sec:model}: that the events are conditionally independent in order to identify unlikely clusters. 

We can use the posterior distribution of the model parameters to estimate the probability of observing $k$ conditionally independent events in some closed hypersphere $b(s_0, r)$ with center $s_0$ and radius $r$. For $\mathbf{X}, \mathbf{Y}$ defined as in Section \ref{sec:model}, we desire 
\begin{align}
 P(d_k(s_0, \mathbf{Y}) \leq r) \label{eqn:prob1}
\end{align}
where $d_k(s_0, \mathbf{Y})$ denotes the distance from $s_0$ to the $k^{th}$ nearest point in $\mathbf{Y}$. \rev{This probability should also be corrected for the problem of multiple comparisons when relevant.} The observed NHPP $\mathbf{Y}$ can only be expressed through the true NHPP $\mathbf{X}$, and ultimately the probability of observing $k$ events within some observed radius will depend on the density of the bounding radius of the unobserved true event positions. Thus the probability in Equation \ref{eqn:prob1} cannot be computed directly given the uncertainties in the true positions of the events. \rev{Otherwise, without noise, this probability could be computed directly using
\begin{align}
    \text{Pr}(d_k(s_0, X) < r) = 1 - \sum_{i=1}^{k-1} \frac{\left(\int_{B(s_0, r)} \Lambda(s) \dd s\right)^i }{i!} e^{-\int_{B(s_0, r)} \Lambda(s) \dd s} \label{eqn:berman},
\end{align}
 reproduced from \citealt{kthnearest}.} Instead, we bound the probability from above using the following:
\begin{multline}
\label{eqn:bound}
P(d_k(s_0, \mathbf{Y}) \leq r) \leq 
    \int_{\boldsymbol{\varepsilon}_1}\dots\int_{\boldsymbol{\varepsilon}_k}\Bigg\{1 - \sum_{i=0}^{k-1}\frac{\left(\int_{B(s_0,\Tilde{r})} \Lambda(s) \mu(\dd s) \right)^i}{i!}\times\\
\exp\left(-\int_{B(s_0,\Tilde{r})} \Lambda(s) \mu(\dd s) \right) \Bigg\} f\left(\{\boldsymbol{\varepsilon}_i\}_{i=1}^k\right)\mu(\dd \boldsymbol{\varepsilon}_1) \dots \mu(\dd\boldsymbol{\varepsilon}_k)
\end{multline} 
 where $\Tilde{r} = r + \max\{|\boldsymbol{\varepsilon}_i|: 1\leq i \leq k\}$,  $f(\{\boldsymbol{\varepsilon}_i\}_{i=1}^k)$ is the joint p.d.f. of $ \{\boldsymbol{\varepsilon_i}\}_{i=1}^k$. Written as the bound of an integral, $\boldsymbol{\varepsilon}_i$ implies the support of $\boldsymbol{\varepsilon}_i$, and $\mu$ is the Lebesgue measure. We derive this bound in the Supplementary Material included for this article. 
 
This bound assumes the most conservative case for an observed radius $r$ and $k$ drawn error displacements: that the largest of those $k$ errors reduced the true (unobserved) $k$-contact radius by the magnitude of the largest error displacement.
 This maximally increases the probability of $k$ events happening in the true NHPP, hence the bound. 
 
 Of course, for all but the most well-behaved $f(\{\varepsilon\}_{i=1}^{k})$ this integral would need to be computed numerically by sampling from $f(\{\varepsilon\}_{i=1}^{k})$. If one can sample from the error distribution, one can sample from the joint distribution of true positions, and hence directly simulate the left hand side of Equation \ref{eqn:bound}. However, the bound in Equation \ref{eqn:bound} has the additional property that, when the $\boldsymbol{\varepsilon}_i$ are i.i.d., it can be further simplified to
\begin{multline}
  P(d_k(s_0, \mathbf{Y}) \leq r) \leq 
     \int_{0}^\infty\Bigg\{1 - \sum_{i=0}^{k-1}\frac{\left(\int_{B(s_0,r+x)} \Lambda(s) \mu(\dd s) \right)^i}{i!}\times \\
    \exp\left(-\int_{B(s_0,r+x)} \Lambda(s) \mu(\dd s) \right) \Bigg\} kf^{k-1}_d(x)\dd x,  
\end{multline}

\noindent where $ kf^{k-1}_{|\boldsymbol{\varepsilon}_i|}(x)$ is the p.d.f. of $\max\{|\boldsymbol{\varepsilon}_i|: 1\leq i \leq k\}$ (also derived in \rev{the Supplement}; \citealt{suppa}). This reduces the computational load by decreasing the number of $d$-dimensional integrations from $k+1$ to two, linearly decreasing the computation time. This would be a large effect when inferences are being made on high-$k$ $k$-contact distances. In our application the $\boldsymbol{\varepsilon}_i$ are not i.i.d. \rev{(although could be approximated as such in very small neighborhoods)} and our $k$ are small, but in Section \ref{sec:rn3} we compute both the bound as well as simulate directly and compare the results. In the following subsection, we explore the utility of the bound in more general applications. 
\subsection{Probability bound simulations} \label{sec:boundsims} To determine how close the newly derived upper bound on $k$-contact probability is to the true $k$-contact distance probability, and hence the utility of the bound, we study two general intensity functions via simulations. Drawing from the first two NHPP intensity models specified by \cite{10.1214/10-ba504}, bivariate Gaussian and Gaussian mixture models, we simulate 50,000 datasets of 2D positions and then perturb all positions from a bivariate normal distribution, $\mathcal{N}_2(0, \sigma_{\text{err}}^2)$. The bivariate Gaussian density is scaled to be an NHPP intensity which integrates to 200 on \rev{its} domain, and with mean vector and covariance matrix
\begin{align}
\mathcal{N}_2 \left( \begin{array}{cc}
0.64  \\
0.61 
\end{array} , \hspace{2mm}
 \begin{array}{cc}
0.016 & 0.007 \\
0.007 & 0.02
\end{array} \right)
\end{align}
respectively. The Gaussian mixture density is also scaled to be a NHPP intensity which integrates to 200 on its domain, with two components, each a bivariate Gaussian with mean vectors and covariance matrices
\begin{align}
q\mathcal{N}_2
 \left( \begin{array}{cc}
0.64  \\
0.61 
\end{array}, \hspace{2mm}
 \begin{array}{cc}
0.016 & 0.007 \\
0.007 & 0.02
\end{array} \right) \hspace{1mm} + \hspace{1mm} (1-q)
\mathcal{N}_2 \left( \begin{array}{cc}
0.25  \\
0.14
\end{array} , \hspace{2mm}
\begin{array}{cc}
0.007 & 0.0005 \\
0.0005 & 0.002
\end{array} \right)
\end{align} 
for the first (mixing weight of $q= 0.71$) and second components respectively.  

For 500 test center points ($s_0$) from these intensities, we then check the frequency in 50,000 simulations that two or more events are found within radius $r=10^{-2}$ of the $s_0$, setting $\sigma_{\text{err}}=10^{-3}$. These values were chosen to explore a range of frequencies from 1.0 to a few/50,000, where 50,000 was chosen to keep simulation time tractable. We then compute the $k$-contact ($k=2$ in this case) probability bound for a radius $r$ for each $s_0$. 

\begin{figure}
    \centering
    \includegraphics[width=0.85\textwidth]{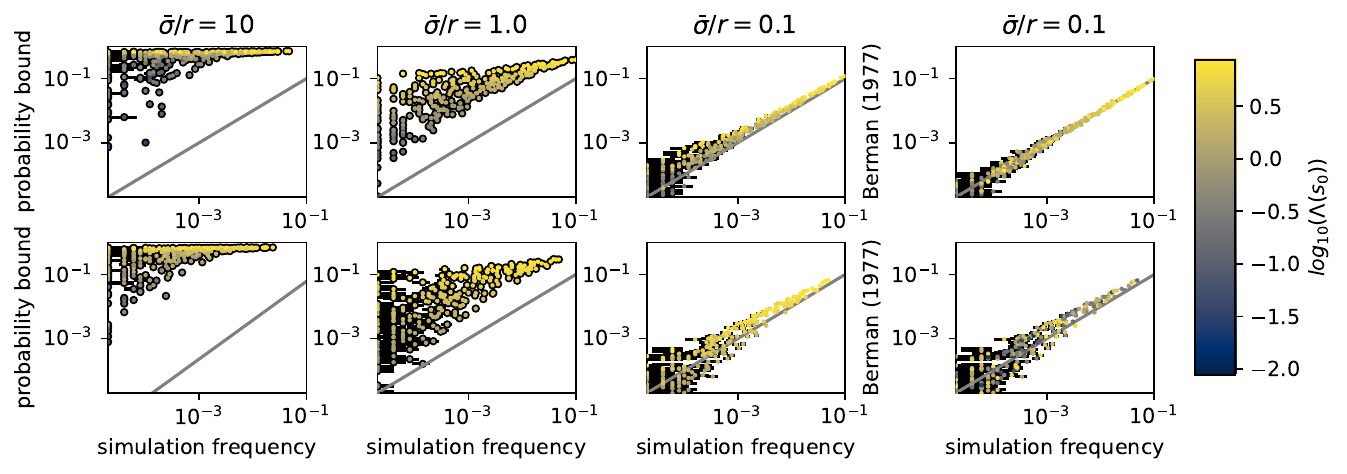}
    \caption{$k$-contact probability bound versus frequency from simulations of two different NHPP intensity models and in different noise regimes. In each panel, 500 test points are drawn uniformly from the domain of the experiment, and then 50,000 simulated datasets are drawn from the relevant NHPP $\Lambda$ and positions are perturbed by Gaussian noise centered at $(0,0)$ and with standard deviation of $10^{-1},  10^{-2},10^{-3}, 10^{-3}$ for the first, second, third, \rev{and fourth columns respectively}. We record the frequency of having $k=2$ events within some radius $r=10^{-2}$ for each simulation. We then, given $s_0$, the noise distribution, and $r$, compute the probability bound in Equation \ref{eqn:bound} \rev{for the first three columns. The final column shows the estimated probability from Equation \ref{eqn:berman}, reproduced from \cite{kthnearest}, ignoring the underlying positional uncertainty in the computation}. Top row: Simulations for test centers ($s_0$) drawn from the bivariate Gaussian intensity (see Section \ref{sec:boundsims}). Bottom row: As for the top row, but simulated datasets are drawn from a bivariate Gaussian mixture intensity. A black line is drawn in each panel to denote where the bound and simulation frequency would be equal. Points are colored according to the value of the intensity at the test center, $\Lambda(s_0)$.} 
    \label{fig:boundsims}
\end{figure}

The simulation frequency vs. probability bound can be found in the leftmost panels of Figure \ref{fig:boundsims}, with points shaded according to the intensity value at $s_0$. The probability bounds range from $0-13$ times larger than the frequencies derived from simulations for both intensity functions in this error and radius regime. We then repeat these simulations for $\sigma_{\text{err}}=10^{-2},$ and $10^{-1}$, such that we can explore the bound in different error regimes: when error displacements are an order of magnitude larger than the radius of interest (\rev{first column}), of comparable size to the radius of interest (\rev{second column}), and an order of magnitude smaller than the radius of interest (\rev{third column}). We find the probability bound is closer to the true value when the test point is at lower intensity regions and when error regimes are small compared to the radii of interest. \rev{Finally, we compute probability in Equation \ref{eqn:berman}, ignoring the underlying error in the smallest tested error regime (fourth column).} 

We conclude from this simulation study that the probability bound is most similar to the true probability for radii similar to or larger than the scale of typical error displacements. If the errors considered are much larger than the $k$-contact distance one wishes to infer, intuitively, the bound is more likely to fail to produce interesting constraints, especially when applied in high intensity regions of the domain. \rev{Also, when we ignore the presence of error, i.e., calculate the probability from Equation \ref{eqn:berman}, which is reproduced from \cite{kthnearest}, at the central test positions for the bivariate Gaussian intensity, we find the estimated probability is an underestimate of the frequency seen in the simulations $5\times$ more often than the bound, which in applications like ours would lead to an increase in type one error. However, this same effect is not observed in the Gaussian mixture intensity, suggesting that while ignoring noise in position measurements may be appropriate in some regimes, it is dependent on factors in addition to error size. Hence, ignoring the effect of error on the probability estimate cannot be deemed universally appropriate.  }
In Section \ref{sec:rn3}, we perform an analysis with the same intention of quantifying the usefulness of the bound for our specific application by comparing the \pcc\, metric from the existing methodology \rev{included in \cite{2023ApJ...947...83C} to that from the methodology proposed in this work, Equation \ref{eqn:bound}.} We show the bound is sufficiently constraining to rule out the hypothesis of $k$ independent sources for more repeater candidates than the existing methodology. 

\section{Application to CHIME/FRB}
\label{sec:intensity}
In our application, the NHPP is a non-uniform intensity of FRB detection probability by CHIME/FRB. The observed FRB detection distribution is a function of the volume density of FRB sources and of the probability of actually detecting an FRB. We first introduce the coordinate system, and then introduce the necessary astrophysical quantities and concepts in order to describe the main factors that affect these distributions and to elucidate how we parameterize the intensity, $\Lambda$. \rev{The motivation for a parametric $\Lambda$ prescription here is that it allows us to leverage our physical intuition and understanding of the instrument to maximize the utility of a relatively  sparse and noisy null dataset.} The application-dependent notation is summarized in Table \ref{tab:astnot}.  
\begin{table}
\centering
\begin{tabular}{cl}
\hline
Symbol & Description \\ \hline
 $\phi$ & Geographical latitude of the telescope\\
 $\delta$ & Declination, equivalent to latitude on the celestial sphere \\
 $\alpha$ & Right ascension, equivalent to longitude on the celestial sphere \\
 DM & Dispersion measure, a rough proxy for FRB distance\\ 
 $\zeta$ & Zenith angle \\
 $N_{\text{FRBs}}$ & Poisson rate, or the integral of the intensity function $\Lambda$ over the entire domain \\
 \hline
\end{tabular}
\caption{\label{tab:astnot} Definition of application-dependent variables used throughout this text.}
\end{table}

\subsection{Coordinate system}\label{sec:coordinates} 
\begin{figure}
    \centering
    \includegraphics[width=0.5\textwidth]{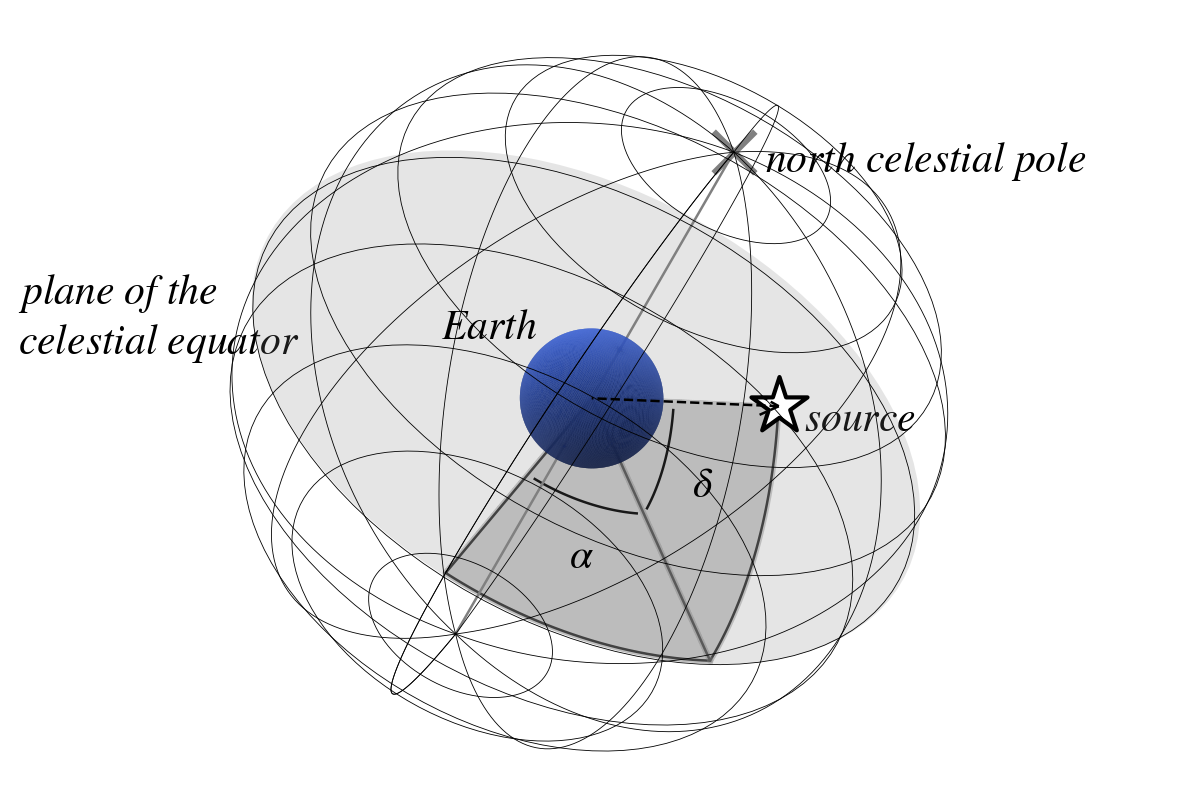}
    \caption{Illustration of right ascension ($\alpha$) and declination ($\delta$). These two angles are equivalent to longitude and latitude on the celestial sphere, respectively. $\alpha$ is referenced from the `vernal equinox', which is one of the intersecting points of the celestial equator and the geographical equator, and $\alpha$ is defined here between $0$ and $360$ degrees. $\delta$ differs from the typical spherical coordinate polar angle as it is referenced to the $x-y$ plane, or the plane of the celestial equator, rather than the $+z$-axis (north celestial pole). $\delta$ is defined between $-90$ (south celestial pole) and $90$ degrees (north celestial pole).  }
    \label{fig:radec}
\end{figure}

FRB positions are defined by right ascension ($\alpha$), declination ($\delta$), and dispersion measure (DM). The angles $\alpha$ and $\delta$ are equivalent to longitude and latitude on the celestial sphere, illustrated in Figure \ref{fig:radec}. 
DM is a quantification of the frequency dependent delay of radio waves due to the ionized gas the waves traverse. As a line-integrated column density, DM is a proxy for distance: if two hypothetical signals from the identical point on the sky were measured to have different DMs, one would necessarily be more distant than the other because the signal with higher DM must have traversed more ionized gas. The common unit for DM of parsecs per cubic centimeter (pc cm$^{-3}$) is used as it is convenient for astrophysicists to determine a rough distance to a Galactic radio source 
given a known average Galactic electron density (in cm$^{-3}$), and has been subsequently adopted for extragalactic radio sources including FRBs. This is partially because the Galactic contribution to the DM of FRBs is considered a foreground contaminant and must be subtracted off in order to make meaningful conclusions about the distance to an FRB using DM. The Galactic contribution comes from ionized gas within the disk-shaped region of our Galaxy (where most stars are located) and the halo of our Galaxy (a much larger gas cloud surrounding this stellar disk). We subtract estimates of these contributions using a model of the density in the disk from \cite{2002astro.ph..7156C} and estimates from \cite{2015MNRAS.451.4277D} of the electron density halo from cosmological simulations which are supported by recent observational results \citep{2023ApJ...946...58C,2023arXiv230101000R}. Henceforth, it is assumed that we are referring to the Galactic-foreground subtracted value when we discuss the DMs of FRBs. 

\subsection{Number of sources vs DM and $\delta$} 
Assuming a Euclidean geometry for the Universe\footnote{This assumption is most appropriate for the very nearby, or low redshift, Universe. The repeater association significance is most complex in the high intensity regions of the NHPP, which happens to be near DM $\sim$ 500 \dmunits\, \citep{2020Natur.581..391M,2022MNRAS.509.4775J,2023arXiv230507022B}, which is a reasonable place to make this assumption.}, and assuming an FRB population that has not changed as the Universe evolved over billions of years,  the source volume number density should be constant as a function of distance from Earth and hence the \rev{cumulative} number of sources within DM, $N_{\text{sources}}$, is found by integrating the volume observed by CHIME
\begin{align}
   N_\text{sources}(\text{DM}, \delta) & \propto \iiint \dd V \propto \int_0^{l} \int_0^{\delta} \int_0^{2\pi} \rho^2 \sin\theta \dd \theta \dd\varphi \dd \rho \propto l^3 \cos(\delta) \\
   & \propto \text{DM}^3 \cos(\delta),
   \end{align}
where $l$ is the distance for which DM is a proxy. \rev{This is similar to the peak DM dependence for FRBs first described by \cite{2018Natur.562..386S}}.  \rev{Explicitly, this is the model for the underlying number of sources, and not the \textit{detectable} number of sources, which we will continue to build up. }

\subsection{Detection probability vs DM and $\delta$} The probability of detecting the source as a function of DM and $\delta$ is complex because it relates to the observational realities of the telescope. We model three main contributing factors: the effects of (\S 4.3.1) the telescope's exposure, (\S 4.3.2) the fluence (defined below) distribution of FRBs, and (\S 4.3.3) the telescope's sensitivity.

\subsubsection{Exposure} The exposure at a declination is the amount of time that the telescope has collected data at that declination. At a fixed $\delta$, CHIME observes some $\alpha$ span instantaneously, which can be described with an arclength on the celestial sphere. This arclength is roughly constant for all $\delta$, forming a long (in the N-S direction) rectangle on the sky. Hence, given some exposure time proportional to the duration of the experiment, where that constant arclength subtends an $\alpha$ angle of 360 degrees (at $\delta$ near 90 degrees), each spot on the sky is seen for the duration of the experiment. For lower $\delta$, this arclength subtends only a fraction of the circumference, and therefore exposure time on a source on that part of the sky is proportional to the fraction of the circumference that the arclength represents. The exposure for each source as a function of $\delta$ is thus approximately proportional to $1/\cos(\delta)$. 
\rev{This approximation breaks down above roughly 70 degrees, where the arclength is no longer small compared to the circumference.} Accounting for this, we model the exposure as a function of $\delta$ as 
 \begin{equation}
     \text{exposure}(\delta) = \frac{c}{1+ d\cos(\delta)}
 \end{equation}
with $c, d$ real positive constants. This function reaches its maximum value of $c$ at $\delta = 90$ degrees and minimum value of $c/(1+d)$ at $\delta = 0$ degrees. This function produces a good fit to the CHIME/FRB exposure data as a function of $\delta$ as released in the most recent catalog of repeater candidates from \cite{2023ApJ...947...83C}.

\subsubsection{Fluence distribution} \label{sec:dm} Fluence is defined as the amount of incident energy on the telescope. Fluence is also hence proportional to the time-integrated energy of the radio burst divided by the distance to the observer squared\rev{, and this affects the DM distribution of FRBs}. The probability of detecting a radio burst falls off with the FRB fluence. For a standard candle or delta-function luminosity distribution\footnote{A `standard candle' in astronomy is a population of sources with a single intrinsic luminosity, and hence whose differing fluences reflect only their distances from Earth.} in Euclidean space, one expects the cumulative fluence distribution to be a power-law with index $-3/2$, $\text{CDF}(F) \propto F^{-3/2}$ where $F$ is the FRB fluence. \rev{Any luminosity function can be approximated by summing arbitratily many such standard candle luminosity functions, and summing arbitrarily many functions with  $\text{CDF}(F) \propto F^{-3/2}$  also produces a function with $\text{CDF}(F) \propto F^{-3/2}$}. 
Simply stated, there is some variation in intrinsic FRB energies, and the farther the source, the smaller the fraction of the distribution of FRB burst energies detectable by the telescope. 

Beyond our descriptions of relative detectable samples as a function of DM and $\delta$, we model the probability of detecting a given source as a negative exponential $(f(F) \propto e^{-F/{F_0}}$ for some threshold $F_0$), as the detection probability is typically modeled as `complete', and hence a steep fall off in probability with decreasing fluence is expected. Complete, in this context, means a burst with a given fluence would be detected with 100\% probability if its position was being observed by the telescope at the time of the burst. The resultant model, \rev{a power law multiplied by a negative exponential,} is reminiscent of canonical galaxy luminosity functions, which compare input luminosities to a characteristic luminosity (we will use a characteristic DM, DM$_0$, which \rev{sets the width of the DM distribution and enforces a peak DM density at DM=$2^{5/3}\text{DM}_0$ in the final intensity below, and then allow for an overall translation parameter DM$_T$ such that the distribution can peak where appropriate)} and are modeled as a negative exponential \citep{1976ApJ...203..297S}. 
\begin{figure}
\tikzset{every picture/.style={line width=0.75pt}} 
\begin{tikzpicture}[scale=0.94,x=0.75pt,y=0.75pt,yscale=-1.5,xscale=1.5]

\draw   (221.2,157.35) .. controls (221.2,157.35) and (221.2,157.35) .. (221.2,157.35) .. controls (221.2,157.35) and (221.2,157.35) .. (221.2,157.35) .. controls (221.2,99.33) and (267.27,52.3) .. (324.1,52.3) .. controls (380.93,52.3) and (427,99.33) .. (427,157.35) -- cycle ;
\draw    (324.1,157.35) -- (324.1,52.3) ;
\draw  [dash pattern={on 4.5pt off 4.5pt}]  (269,68.5) -- (324.36,156.89) ;
\draw   (263,53.4) -- (264.73,57.21) -- (268.61,57.82) -- (265.81,60.79) -- (266.47,64.98) -- (263,63) -- (259.53,64.98) -- (260.19,60.79) -- (257.39,57.82) -- (261.27,57.21) -- cycle ;
\draw  [draw opacity=0] (334.63,138.23) .. controls (340.25,142.36) and (343.96,149.52) .. (343.96,157.66) -- (323.54,157.66) -- cycle ; \draw   (334.63,138.23) .. controls (340.25,142.36) and (343.96,149.52) .. (343.96,157.66) ;  
\draw  [draw opacity=0] (310.11,134.21) .. controls (313.98,131.44) and (318.59,129.83) .. (323.54,129.83) -- (323.54,157.21) -- cycle ; \draw   (310.11,134.21) .. controls (313.98,131.44) and (318.59,129.83) .. (323.54,129.83) ;  
\draw    (455,172) -- (485,172) ;
\draw [shift={(485.87,172)}, rotate = 180] [fill={rgb, 255:red, 0; green, 0; blue, 0 }  ][line width=0.08]  [draw opacity=0] (5.36,-2.57) -- (0,0) -- (5.36,2.57) -- cycle    ;
\draw [shift={(452,172)}, rotate = 0] [fill={rgb, 255:red, 0; green, 0; blue, 0 }  ][line width=0.08]  [draw opacity=0] (5.36,-2.57) -- (0,0) -- (5.36,2.57) -- cycle    ;
\draw    (338,173) -- (326,161) ;
\draw [shift={(325,160)}, rotate = 42] [fill={rgb, 255:red, 0; green, 0; blue, 0 }  ][line width=0.08]  [draw opacity=0] (6,-2) -- (0,0) -- (6,2) -- cycle    ;
\draw  [color={rgb, 255:red, 155; green, 155; blue, 155 }  ,draw opacity=1 ] (375.52,66.75) -- (324.36,156.89) -- (233.85,105.52) ;

\draw  [draw opacity=0] (306.09,146.42) .. controls (307.61,143.21) and (309.78,140.46) .. (312.39,138.39) -- (324.1,157.35) -- cycle ; \draw   (306.09,146.42) .. controls (307.61,143.21) and (309.78,140.46) .. (312.39,138.39) ;  
\draw  [color={rgb, 255:red, 155; green, 155; blue, 155 }  ,draw opacity=1 ] (315.42,151.81) -- (320.33,143.17) -- (329.27,148.25) ;
\draw (378,58) node [anchor=north west][inner sep=0.75pt]   [align=left] {\textit{north celestial pole}};
\draw (310,42) node [anchor=north west][inner sep=0.75pt]   [align=left] {\begin{minipage}[lt]{29.94pt}\setlength\topsep{0pt}
\begin{center}
\textit{zenith}
\end{center}
\end{minipage}};
\draw (296,130) node [anchor=north west][inner sep=0.75pt]   [align=left] {$\displaystyle \delta $};
\draw (344,138) node [anchor=north west][inner sep=0.75pt]  [font=\small] [align=left] {$\phi$};
\draw (310,116) node [anchor=north west][inner sep=0.75pt]   [align=left] {$\zeta$};
\draw (338,170) node [anchor=north west][inner sep=0.75pt]   [align=left] {\textit{telescope}};
\draw (492,173) node [anchor=mid][inner sep=0.75pt]  [font=\small] [align=left] {N)};
\draw (447, 173) node [anchor=mid][inner sep=0.75pt]  [font=\small] [align=left] {(S};
\draw (234,51) node [anchor=north west][inner sep=0.75pt]   [align=left] {\textit{source}};
\draw (430,151) node [anchor=north west][inner sep=0.75pt]  [font=\small] [align=left] {\textit{ground/horizon}};
\draw (172,97) node [anchor=north west][inner sep=0.75pt]   [align=left] {\textit{celestial equator}};
\draw (374.0,60.48) node [anchor=north west][inner sep=0.75pt]  [font=\small,color={rgb, 255:red, 155; green, 155; blue, 155 }  ,opacity=1 ,rotate=-27.85][align=left] {$\displaystyle \times $};

\draw (228,106.8) node [anchor=north west][inner sep=0.75pt]  [font=\small,color={rgb, 255:red, 155; green, 155; blue, 155 }  ,opacity=1 ,rotate=-307.9][align=left] {$\displaystyle \times $};

\end{tikzpicture}
\caption{Illustration depicting the horizon-to-horizon view of the CHIME telescope to show the zenith angle ($\zeta$) of a given source (indicated by a star) as a function of the declination of that source ($\delta$). Zenith is defined as perpendicular to the horizon. The north celestial pole and celestial equator are defined as perpendicular to one another, and the north celestial pole is the axis around which the Earth spins. The angle between the north celestial pole and the horizon is equal to the geographic latitude ($\phi$) of the telescope \citep[$\phi=49.32$ degrees;][]{2022ApJS..261...29C}. The declination is defined as the angle between the source and the celestial equator. Hence we can calculate the zenith angle, $\zeta(\delta)$, of a source as $\zeta = \phi - \delta$.} 
\label{fig:zeta}
\end{figure}
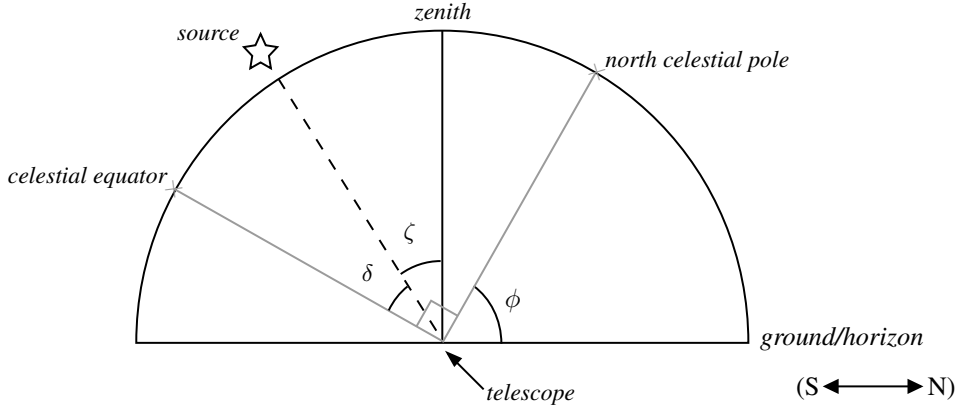

\subsubsection{Sensitivity} The sensitivity of a telescope describes the minimal brightness that the telescope can distinguish from random noise. The sensitivity of the CHIME telescope varies with location on the sky; CHIME is most sensitive directly overhead, at the telescope's zenith, and the sensitivity decreases as a function of the zenith angle $\zeta$ (Figure \ref{fig:zeta}). 
Increased sensitivity means that a larger fraction of the luminosity function can be detected at a given distance, and thus one expects the DM with maximal probability of being detected to decrease as a function of increasing $\zeta$. Therefore we modulate the term representing the maximal density DM, $\text{DM}_0$, by a factor of $1+\cos(\zeta(\delta)) = 1+\cos(\phi - \delta)$ where $\phi$ is the geographic latitude of the telescope. \rev{The addition of one prevents zero values in the denominator of the intensity.} 

The global effect on the sensitivity as a function of $\delta$ can be described by $\cos^b(\zeta(\delta))$, where $b>0$ is a constant, introduced to fit the steepness of the drop off in sensitivity with $\zeta$. However, the way that CHIME/FRB forms beams to search for FRBs on the sky means the sensitivity fluctuates around $\cos^b(\zeta)$ as a function of $\delta$, with some positions falling in gaps between the most sensitive areas of each beam. Integrating over all observing frequencies and $\alpha$, this effect is small compared to the sensitivity changes as a function of distance away from zenith, and median sensitivity is still well-described by $\cos^{b}(\zeta)$. Thus we do not model the small-$\delta$-scale sensitivity behavior in the global intensity function, and instead rely on its imprint on the individual burst localization uncertainty regions, discussed in Section \ref{sec:noise}, to capture its contribution to the uncertainty in the posterior distributions. While we can make a reasonable estimate of $b$ based on our beam model,\footnote{available at \url{https://chime-frb-open-data.github.io/beam-model/}} the exact value depends on how CHIME/FRB forms beams on the sky and the spectra and polarization of the observed FRBs, therefore we fit $b$ rather than assert it in the intensity function \citep{2015A&A...576A..62N, overview}. 

\begin{figure}
    \centering
    \includegraphics[width=0.85\textwidth]{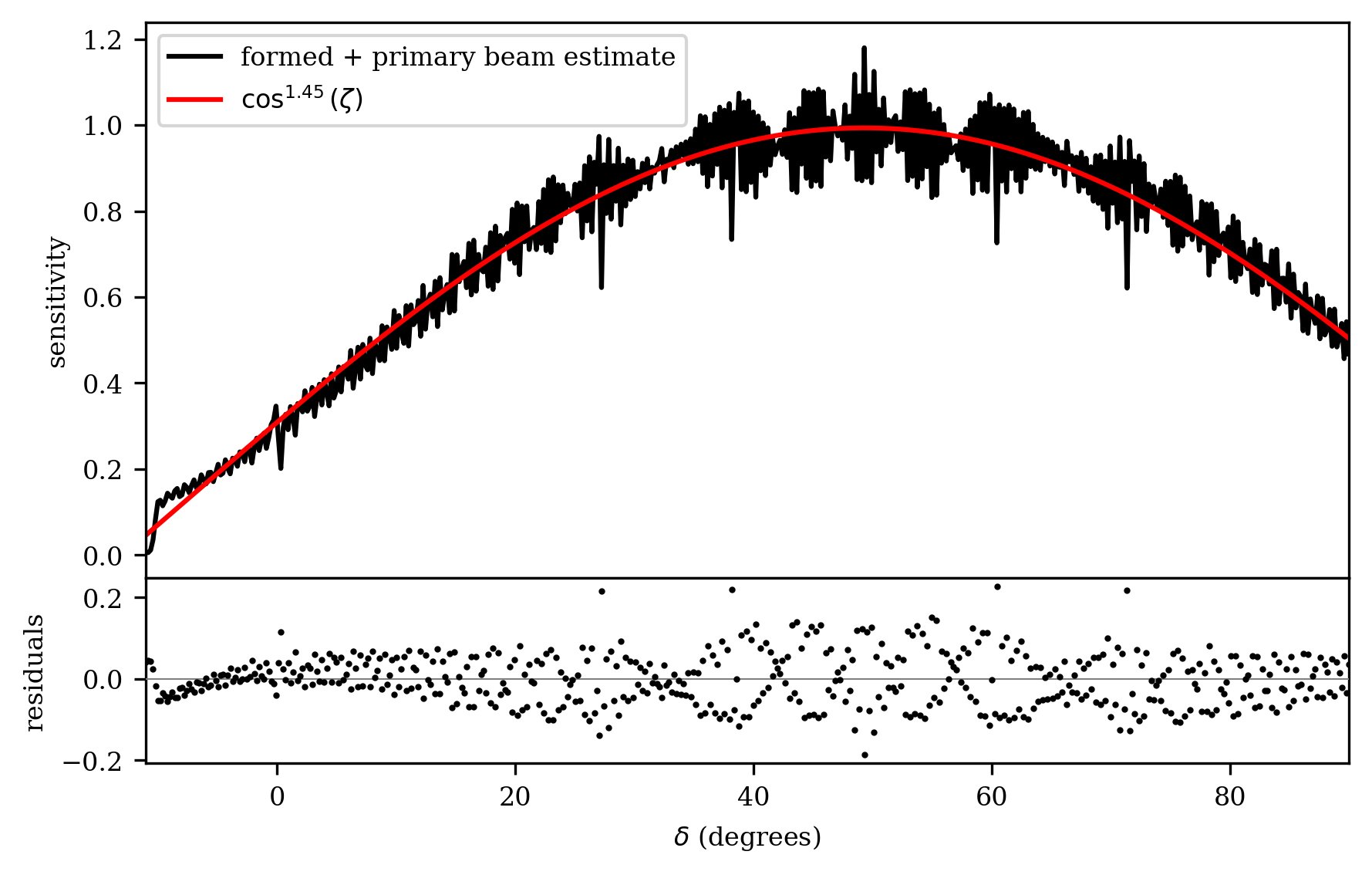}
    \caption{\textit{top panel}: Sensitivity as a function of declination and normalized to attain a max value at 1 at our zenith, intended to compare the behaviour of our best sensitivity estimate and $\cos^b(\zeta)$ (black line). The sensitivity estimate was modeled using CHIME/FRB's publicly available beam-model, available at \url{https://chime-frb-open-data.github.io/beam-model/}
     by summing the two components of CHIME/FRB's beam, the `formed' beam and the `primary' beam (a detailed description is more technical than appropriate to be included here, but can be found in \citealt{overview}). The sensitivity is expected to follow $\cos^b(\zeta)$ where $b$ is a constant which depends on how CHIME/FRB forms beams on the sky as well as the polarization of FRBs \citep{2015A&A...576A..62N, overview}. We use a simple least squares fit in order to estimate a value for $b$ to show that the median of the sensitivity is well described by such a function. Bottom panel: residual values.} 
    \label{fig:sensitivity}
\end{figure}

We show the empirically estimated model of the telescope's sensitivity as a function of $\delta$ in Figure \ref{fig:sensitivity} with a least squares fit to our model for this component, $\cos^b(\zeta(\delta))$ for illustrative purposes. There is a slight disagreement between the best-fit value $b=1.45$ and the plotted sensitivity model, especially at low $\delta$, but this disagreement will be reflected in the posterior distribution of $b$ and occurs at a low-intensity region where the repeater significance is least ambiguous.

The intensity function we use thus defined 

\begin{multline}\label{eqn:int}
 \Lambda (\delta, \text{DM}) | N_{\text{FRBs}}, b, c, d, \text{DM}_T,\text{DM}_0 \propto N_{\text{FRBs}} \exp \Bigg( \frac{c}{1 + d\cos(\delta)} - \\ \left( \frac{\text{DM}- \text{DM}_T}{\text{DM}_0(1+\cos^b(\zeta))} \right)^{3/2}  \bigg) \cos(\delta)
 \left( \frac{\text{DM}- \text{DM}_T}{\text{DM}_0(1+\cos^b(\zeta))} \right)^3.
\end{multline} 
where we use $\propto$ rather than an equality because the total intensity should integrate to the expected rate of FRBs, $N_{\text{FRBs}}$. 

 \subsection{Measurement Error}
\label{sec:noise}
\begin{figure}
    \centering
    \includegraphics[width=0.8\textwidth]{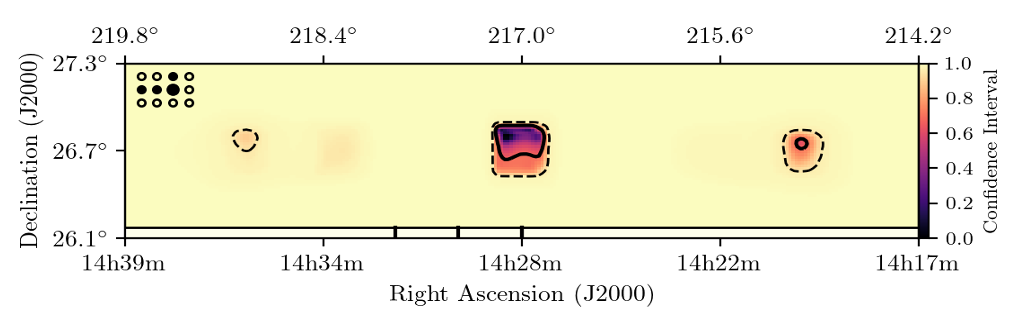}
    \caption{An example of uncertainty regions for FRB localizations with CHIME/FRB's real-time pipeline, to demonstrate the multi-modal nature. The colormap shows which confidence interval percentage one would need to consider in order to include that position in the localization error. 
    In this plot, the unit for $\alpha$ is not fractional degrees but rather hours and minutes of arc, where 360 degrees are equal to 24 archours, and one degree is 60 arcminutes. The dots in the upper left corner illustrate the pattern of CHIME/FRB's beams in which the burst was detected. Credit: \cite{2021ApJS..257...59C}. }
    \label{fig:locerror}
\end{figure}

Each CHIME/FRB burst detection has a unique localization p.d.f. (which describes the noise or measurement error) due to its beam configuration, described in detail in Section 3.1 of \cite{2021ApJS..257...59C}. CHIME/FRB's real-time pipeline calculates the most likely positions on the sky for detected bursts using ratios among per beam signal-to-noise values fit 
to beam model predictions for a grid of model sky locations and model intrinsic spectra. The confidence
interval map or distribution is estimated from a sample of events from radio transients with precisely known positions also detected by CHIME/FRB, such that true positions fall within contours of a given confidence interval with the appropriate frequency. CHIME/FRB's primary beam sensitivity in the E–W direction, mostly concerning the $\alpha$ of sources, has roughly the form of a $\sinc(x) \equiv \sin(x)/x $ function with origin at the meridian (directly overhead the telescope in the $\alpha$ plane) of the telescope. The localization uncertainty regions reported are chosen to contain the `main lobe' (or largest amplitude oscillation, using the sinc function analogy) of the primary beam and the first order `side lobes' (the oscillations to the left and right of the largest amplitude of the sinc function) of the formed beams, which leads to multi-modal uncertainty regions. An example of these uncertainty regions is shown in Figure \ref{fig:locerror}. We sample true positions from the uncertainty regions of each burst when estimating the posterior distributions of the intensity function.

\subsection{Putting it all together}
\label{sec:implementation}
\paragraph*{Hyperprior distribution} Our Bayesian approach to model fitting requires the specification of a prior distribution. We assume prior independence among the hyperparameters and specify the joint hyperprior distribution as the product of the following univariate distributions: 
\begin{align*}
    N_{\text{FRBs}} & \sim \text{Uniform}(128.8,2362.8) &     \text{DM}_0 & \sim\mathcal{N}\left(127.8,127.8\right) & \text{DM}_T & \sim\mathcal{N}(156,156)\\ 
    b & \sim\mathcal{N} (1.45, 0.12)
  & c & \sim \text{Uniform}(0, 10)
      & d & \sim \text{Uniform}(0, 10).
\end{align*}
We elicit the prior distribution on $N_{\text{FRBs}}$  based on the known survey duration of 214.8 days \citep[data not otherwise used in this work;][]{2021ApJS..257...59C} and a pre-experiment estimate of CHIME/FRB's detection rate of 0.6-11 bursts/day \citep{overview}, which was calculated using the all-sky FRB rate from \cite{2017AJ....154..117L} and adjusted for CHIME/FRB's survey design by \cite{2017ApJ...844..140C}. $N_{\text{FRBs}}$ can be any real number rather than only natural numbers as it is treated as the Poisson rate over the domain, where the likelihood in Equation \ref{eqn:likelihood} is computed as the likelihood of $n\in\mathbb{N}$ events in our sample given the overall rate of $N_{\text{FRBs}} \in \mathbb{R}$. The center and standard deviation of the Normal prior distribution of $b$ is based on the best fit value and least squares error of the fit to the beam model. Our distribution of $\text{DM}_0$ and $\text{DM}_T$ are based on detected FRBs from surveys other than CHIME/FRB: we download all reported FRBs on the transient name server\footnote{\url{https://www.wis-tns.org/} accessed Aug 1st, 2024} and exclude those detected by CHIME/FRB (our data). We then take the sample average and standard deviation of these non-CHIME/FRB FRB \rev{extragalactic} DMs, approximately 560 and 404 \dmunits \, respectively. \rev{These values are subtracted from eachother for the translation term DM$_T$ and scaled by $2^{-5/3}$ to reach the corresponding peak distribution value for DM$_0$ (see Section \ref{sec:dm}).}



\paragraph*{Computational Specifics} We now take the Bayesian approach to fitting the model specified in Section \ref{sec:model} for the functional form of our application's intensity function. \rev{We fit our model to a well-studied subset of FRBs, CHIME/FRBs first catalog \citep{2021ApJS..257...59C}, using a bespoke Metropolis-Hasting within block Gibbs sampler to explore the joint posterior distribution \citep[see, e.g.,][for a description of such samplers using astrophysical examples]{2018sabm.book...29S}}. 
Our sampling algorithm is described in Algorithm \ref{alg:cap}.

\begin{algorithm}
\caption{MCMC sampling for generalized Hierarchical Bayesian Spatial Process}\label{alg:cap}
\begin{algorithmic}[1]
\For{$i = 1 \to n$ samples:}
    \State block update from multivariate normal distribution of hyperparameters with Metropolis–Hastings (MH)
    \For{$j = 1 \to $len(training data):}
    \State block update for event positions ($\alpha_j, \delta_j$) with MH independence sampler
    \EndFor
    \For{$j = 1 \to $len(training data):}
    \State single-site update for burst DM with Metropolis sampler
    \EndFor
\EndFor 
\State\textbf{Iterate}
\end{algorithmic}
\end{algorithm}

Leveraging the posterior distributions of the hyperparameters, we can then make a predictive inference on the $k$-contact distance at a given $\alpha, \delta,$ DM for the current CHIME/FRB FRB detection count. We can quantify the probability of detecting $k$ or more bursts within the observed distance of the minimal bounding sphere ($b(s_0, r) \subset D)$ or closer, under the null hypothesis that all bursts are physically independent. 

 To propose new values during each iteration of our MCMC routine, we use the following jumping distributions 
\begin{align*}
(N_\text{FRBs,i+1},b_\text{i+1},c_\text{i+1},d_\text{i+1}, \text{DM}_{0 ,\text{i+1}}, \text{DM}_\text{T,i+1})^{\text{T}}\hspace{-1.5mm} \sim \mathcal{N}_6 ( ( N_\text{FRBs,i},b_\text{i},c_\text{i},d_\text{i}, \text{DM}_\text{0,i},\text{DM}_\text{T,i})^{\text{T}}, \boldsymbol{\mathbb{M}}_{ij}),
\end{align*}
with a $6\times6$ real-valued, positive-semi-definite covariance matrix $\boldsymbol{\mathbb{M}}_{ij}$ estimated from the first 1000 iterations of an MCMC run, which are discarded as burn in and after which $\boldsymbol{\mathbb{M}}_{ij}$ remains fixed. For the block updates of $\alpha_i, \delta_i , \ i \in {1,2, \ldots, 527}$, we use an independence sampler. This is a version of a Metropolis-Hastings sampler wherein the proposal distribution is random within the domain, weighted by an approximation of the posterior distribution (in our case, $\boldsymbol{\varepsilon}_i)$, and hence depending on the previous iteration only in the acceptance step of the algorithm. We weight according to an empirical estimate of the p.d.f. for the position of each burst (more information in Section \ref{sec:noise}). We do single-site Metropolis-Hastings updates for the DM of each burst using a Normal jumping distribution centered at the previous iteration value of the DM and with standard deviation equal to the estimated measurement standard deviation, typically between 0.4 to 3 \dmunits, reflecting the telescope's real-time pipeline precision.

\section{Performance}
\label{sec:performance}
\subsection{Convergence}
In addition to visual inspection of the chains and acceptance rate monitoring, we diagnose convergence using ten different chains with starting values chosen via latin hypercube sampling \citep{ef76b040-2f28-37ba-b0c4-02ed99573416} from the hyperprior. This tests if the results are insensitive to the choice of starting values for the MCMC routine. 
Visual inspection of the ten trace plots suggests that each chain converged to the same apparent stationary distribution. 
The Gelman and Rubin $\hat{R}$ statistic \citep{10.1214/ss/1177011136} of each of the hyperparameters is no greater than 1.004, which does not indicate a lack of convergence. In addition, we compute the effective sample size of the  chains using the \texttt{ArviZ} package\footnote{version 1.2.0, \url{https://github.com/arviz-devs/arviz}} \citep{2020ascl.soft04012A} and find no value smaller than \rev{1183}, which is sufficient for our purposes.
\subsection{Simulation Study of Intensity function hyperparameters}
In order to test the reliability of this methodology, we test an NHPP described by our intensity function with known hyperparameters, chosen to be $N_{\text{FRBs}}, b, c, d, \text{DM}_0, \text{DM}_T$ equal to 525, 1.5, 6, 2, 560, 400 respectively. These values are chosen as they are similar to the median hyperparameter values derived for our data. We randomly draw ten samples of events from this known intensity function by (i) drawing the number of observed events, with a Poisson distribution with rate equal to 525, (ii) selecting that number of positions from the known $\Lambda$ using rejection sampling, and then (iii) perturbing these positions to simulate the noisy observations. We then run our MCMC method, outlined in Section \ref{sec:implementation}, on the output noisy datasets. 
The chains are started at parameter values drawn from their hyperpriors using the same latin hypercube sampling technique as before: we divide the interval (0,1) into 10 (number of chains) subintervals, randomly sample a value from those subintervals corresponding to each hyperparameter, and then draw the value of the starting hyperparameter from the \rev{inverse} c.d.f. of its marginal hyperprior. We then randomly permute these starting parameter values, without replacement. 
Since the covariance matrix we used to propose new values for the hyperparameters was tuned to the real dataset, we allow the chains to update the jumping distribution matrix $M_{ij}$ to the chain's current covariance at iteration 1000 and remove all iterations before this as burn-in. We compare the posterior distributions to the known values and find that the true value of the $N_{\text{FRBs}}, b, c, d, \text{DM}_0, \text{DM}_T$ hyperparameter falls within the 90\% credible interval of the estimated posterior distribution for 10, 10, 9, 8, 10, 10 simulations out of 10 total simulations, respectively, indicating that the method has good frequentist coverage for our intensity function.




\section{Results}
\label{sec:results}
\subsection{Posterior distribution of the intensity hyperparameters}
\begin{figure}
    \centering
    \includegraphics[width=\textwidth]{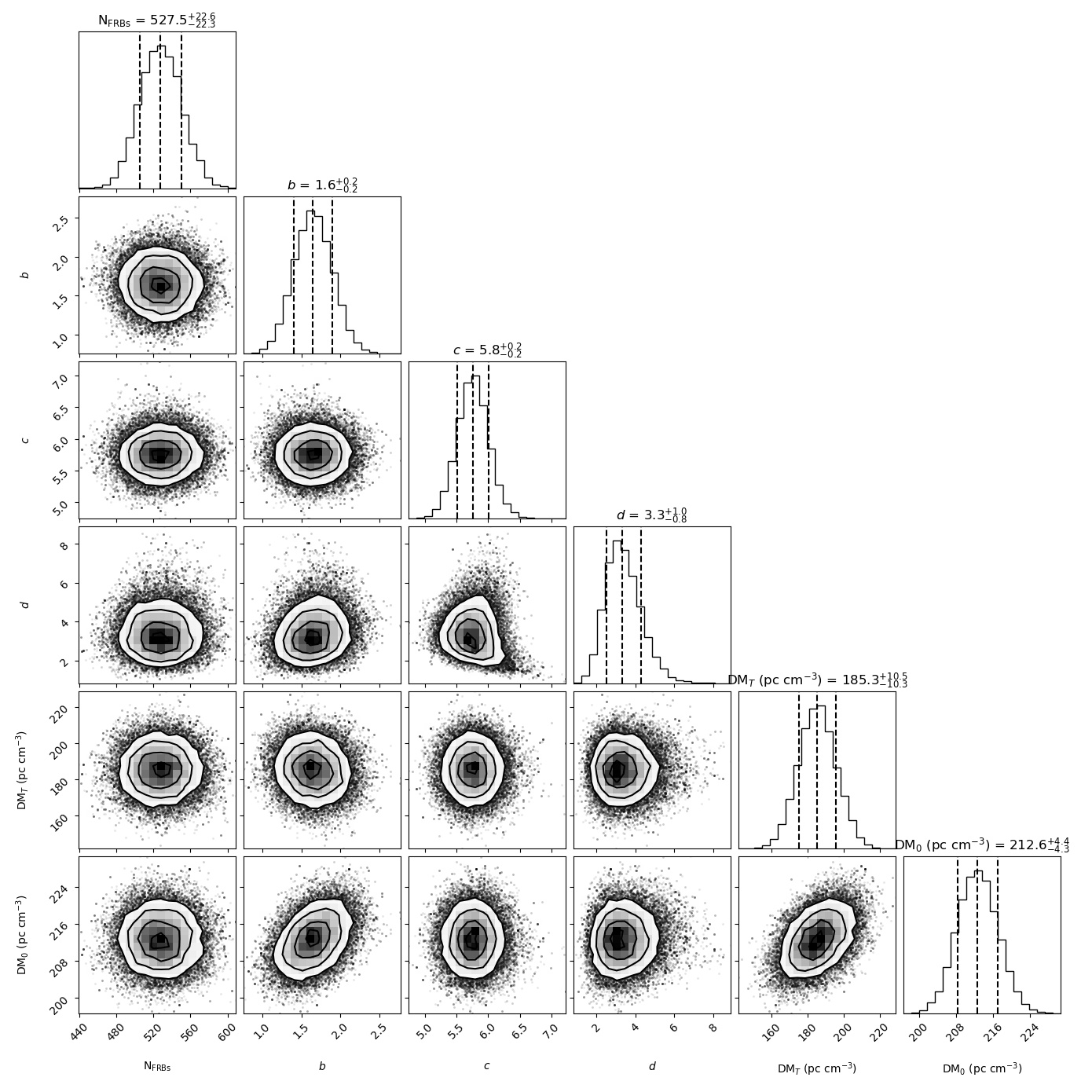}
    \caption{\label{fig:corner}Corner plot of non-homogeneous spatial Poisson intensity hyperparameters from the (approximate) posterior draws obtained with MCMC. The values written in the titles of the histograms correspond to the median posterior distribution estimate for each parameter, and the errors represent the \rev{width} of their 68\% credible regions. }
\end{figure}

\begin{figure}
    \centering
    \includegraphics[width=0.9\textwidth]{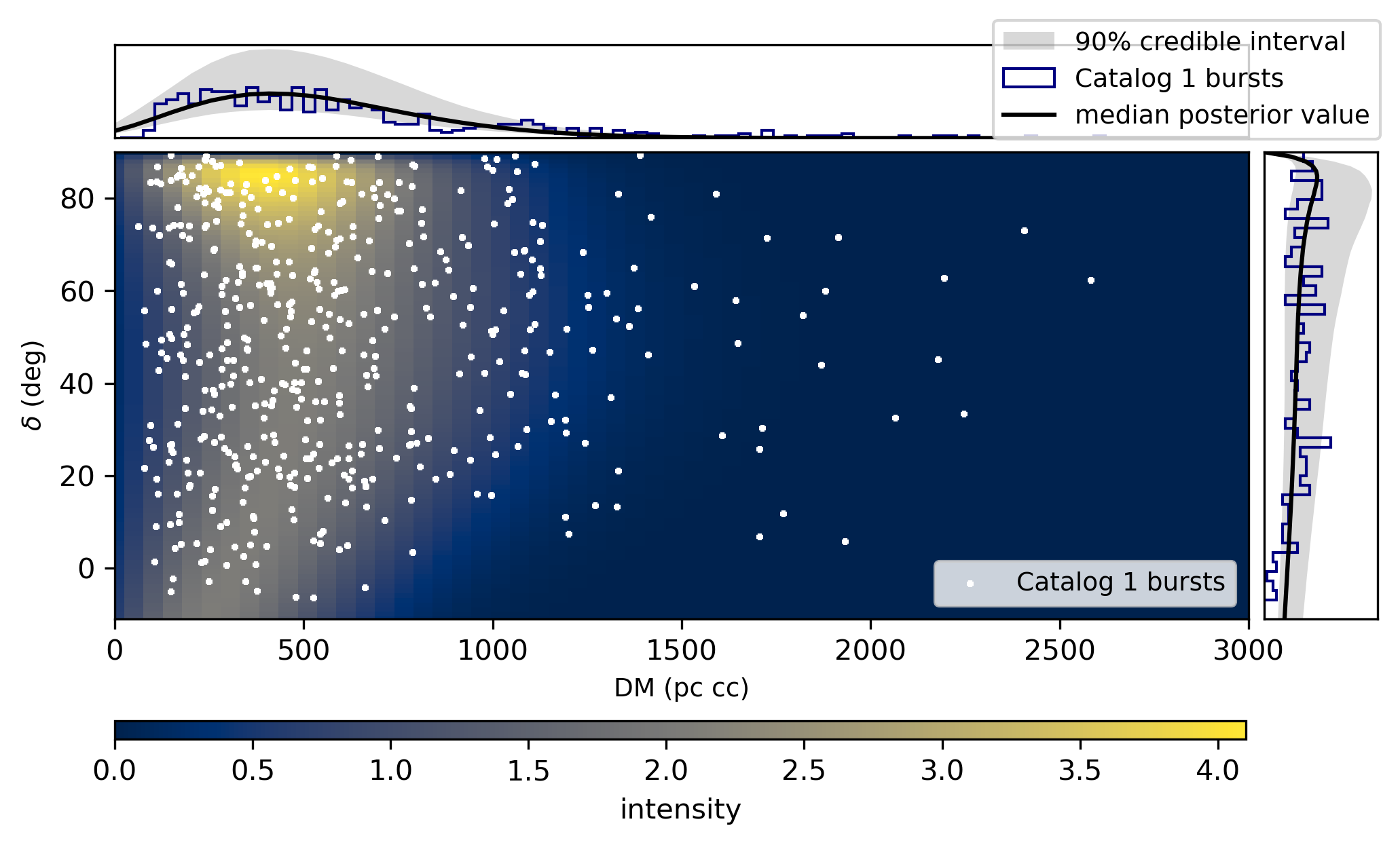}
    \caption{\label{fig:intensity}Center plot: CHIME/FRB intensity predicted from the median posterior hyperparameters. We overplot the FRB detections in the First CHIME/FRB catalog \citep{2021ApJS..257...59C}, our training set, as a function of $\delta$ and DM (white points). In the right and top subplots, we show a normalized histogram of these FRB detections (navy)  as a function of $\delta$ and DM, respectively. In these panels, we also show the 90\% pointwise credible interval (grey regions) and median point-wise value (black lines) from the posterior distribution.}

\end{figure}

In Figure \ref{fig:corner} we display a corner plot of the MCMC samples. Using the sample median of the draws for each hyperparameter as an estimate of the posterior median, we show the corresponding median intensity in Figure \ref{fig:intensity} for illustrative purposes. In the marginals, we show the pointwise 90\% credible region from all draws from the posterior distribution, as well as the pointwise median. We overlay the null data set detections to compare the realization of the intensity in the catalog/null dataset. While this intensity form is required in order to compute our motivating statistic, \pcc, it also confirms that the instrument's intensity function is well-fit by a model based on known physics and spherical geometry. \rev{We also perform a residual-based model evaluation, motivated by the procedure of \cite{https://doi.org/10.1111/j.1467-9868.2005.00519.x}, which does not suggest significant model misspecification or discrepancy; see the Supplement \citep{suppa} for details.}


\subsection{\cite{2023ApJ...947...83C} Repeater sample}
\label{sec:rn3}
We estimate the \pcc\, value for each of the repeater candidates of \cite{2023ApJ...947...83C} both using the bound in Equation \ref{eqn:bound} and directly from simulations. The simulations sample true/unobserved positions from estimates of their error density distributions and then compute the probability in Equation \ref{eqn:berman}. The bound in Equation \ref{eqn:bound} is an upper limit on the probability of a $k$-contact distance smaller than or equal to the observed value for each of the repeater candidates.

Both the probability and probability bound estimations require integration which is computed via Monte Carlo. Each random draw for the Monte Carlo integration uses a set of hyperparameters from our posterior distribution for $\Lambda$, but with $N_{\text{FRBs}}$ multiplied by a factor which reflects the increased number of events in the training set versus observed events in the \cite{2023ApJ...947...83C} dataset (i.e., 2196/527, the ratio of number of observed events in the repeater catalog to the number of observed events in the training data). The observed value of the $k$-contact distance and the repeater's observed position used in these methods are defined by estimating the minimal bounding sphere of the candidate repeater.
 We estimate the minimal bounding sphere using the algorithm described in Chapter V.2 of \cite{glassner1990graphics}.

\begin{figure}
    \centering
    \includegraphics[width=0.85\textwidth]{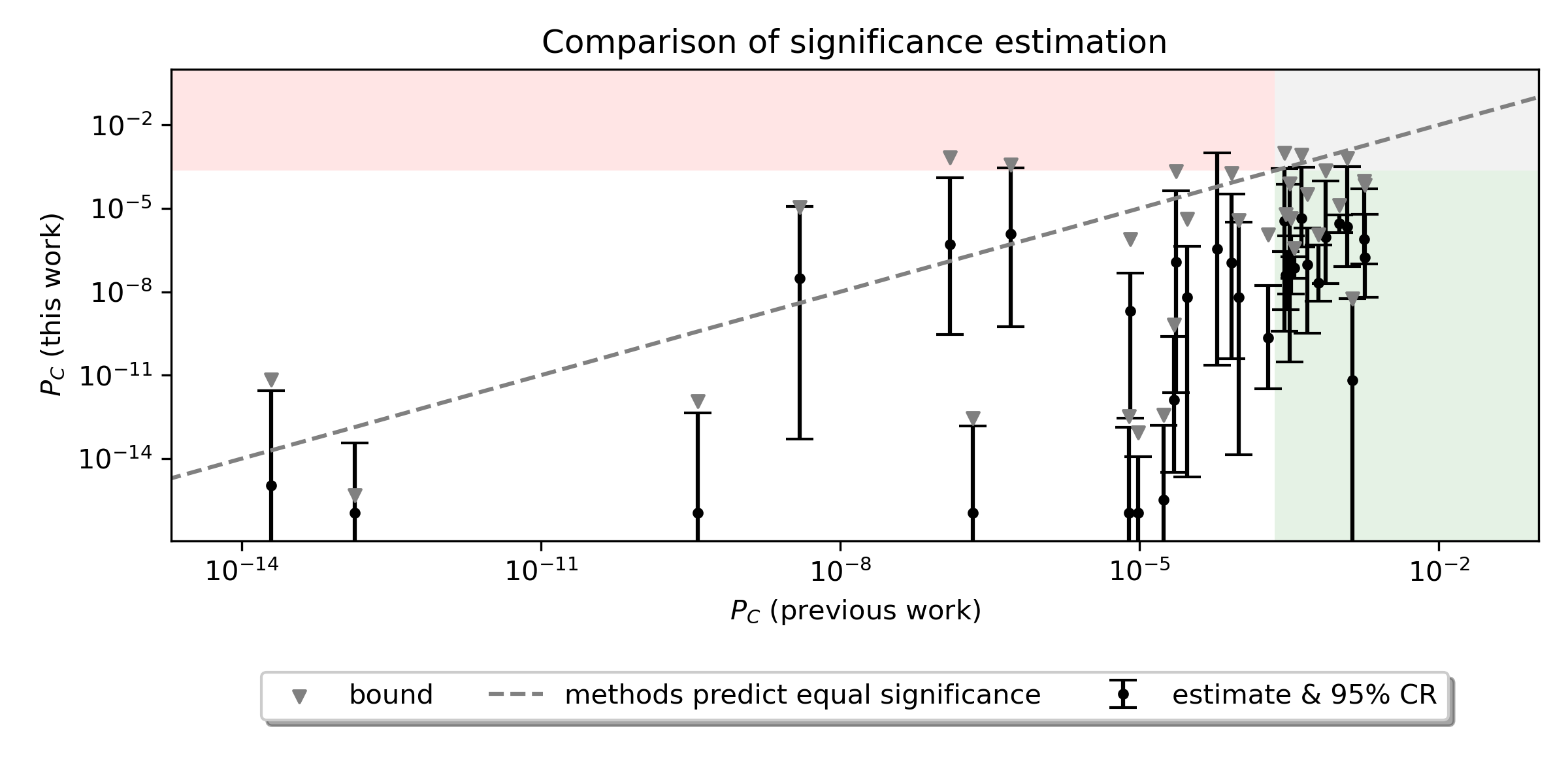}
    \caption{Comparison of \pccl\, (\pcc) 
    from the previous method ($x$-axis) and from the method derived in this work ($y$-axis) for repeater candidates from \cite{2023ApJ...947...83C}. This work presents both a probability estimate from simulations (black points, errorbars represent the 95\% credible region) and a probability bound estimate (grey triangles). We denote the line at which the estimates from the new and old methods would be equal with a gray dotted line. Candidates in the unshaded plot region would be classified as having sufficient evidence to call the source a repeater in both methods, and candidates in the grey shaded region have \pcc\, estimates from both methods that would suggest the candidates are ambiguous. The thirteen candidates in the green region are candidates that can be newly declared unambiguous repeater candidates using the \pcc\, method outlined in this paper. } 
    \label{fig:rn3pcc}
\end{figure}

In Figure \ref{fig:rn3pcc} the values are plotted against a modified value of the existing statistic for these sources, $R_{\text{CC}}$. $R_{\text{CC}}$ estimates the false-positive contamination rate within the repeater sample\footnote{ \rev{That is, \cite{2023ApJ...947...83C} present a \pcc\, statistic which is corrected for multiple-hypothesis testing. This correction is necessary due to the large number of FRBs considered. Detailed discussion of candidate identification and the implied trials factor can be found in \cite{2023ApJ...947...83C}.}} assuming each FRB detection is an independent binomial trial, where a `success' is a detection within some box around the repeater candidate, and a `failure' is a detection outside. We multiply these $R_{\text{CC}}$ values by the trials factor such that the values can be considered a \pccl\, and can be directly compared with our \pcc\, estimates, and plot this value  $R_{\text{CC}}/N$ denoted \pcc\, (previous work). The intensity function of \cite{2023ApJ...947...83C} is estimated by smoothing the two dimensional detection histogram with a Gaussian kernel, and the probability of success/failure is approximated as the fraction of intensity inside/outside of the box. The main issues with this method are the ambiguity of defining the `in' box, and that it does not take into account measurement error other than the 95\% span of the localization and DM errors, inside of which all possibilities are equally weighted despite the multi-modal nature of the error regions. The intensity function also doesn't properly account for measurement error, and the choice of Gaussian kernel could potentially obscure minute fluctuations in intensity, or could overweight a stochastic over- or under- density. 

In \rev{31 of 39 cases}, our new methodology returns a smaller median \pcc\, for the repeater candidate, meaning the candidates are considered less likely to be independent sources, with a median improvement factor (original \pcc/ our \pcc\, estimate) of \rev{about 4800}. This includes \rev{14} candidates that were previously labeled as ambiguous due to their high \pcc, whereas the new methodology produces values of \pcc\, that suggest that the sources are unambiguously repeaters (assuming the same $p$-value threshold as adapted by \citealt{2023ApJ...947...83C}). \rev{    In three of the remaining cases, the median probability from our new methodology is slightly above the previous estimate. The only common feature between them seems to be their DMs, which are all within the relatively common DM range of $\sim$ 600--750. It is possible that the intensity was slightly underestimated at these DMs in the old methodology and hence the general increase in sensitivity we are seeing with this methodology is diminished by the increased intensity in the region.} In the final \rev{five} cases, the returned estimated probability is zero, reflecting that the estimate is smaller than can be accurately captured by our double-precision floating-point format (roughly $10^{-16}$). \rev{This conclusion is supported by these clusters each having between 5--12 events, which are among the highest cluster sizes and extremely unlikely to happen by chance.} 

Our Bayesian methodology allows for uncertainty quantification in the estimate. In the direct simulation of \pcc\, we randomly sample true positions from the localization distributions and randomly sample $\boldsymbol{\theta} = \{N_{\text{FRBs}}, b, c, d, \text{DM}_0, \text{DM}_T\} $ for $\Lambda$ from our estimate of its posterior distribution. We then compute the probability in Equation \ref{eqn:berman} assuming these true positions and this set of hyperparameters and repeat 5000 times. The 95\% credible regions on the value of \pcc\, from this method are also shown in Figure \ref{fig:rn3pcc}. \rev{The sizes of these regions seem to be mostly informed by the uncertainty in the measured positions}. Considering the credible regions, it is clear that the \pcc\, estimates tend to be lower than the \pcc\, (previous work) value. For the repeater candidates whose median simulated value of \pcc\, are lower than the threshold $p$-value in \cite{2023ApJ...947...83C}, the span of the 95\% credible region tends to be entirely below this threshold $p$-value. The width of the 95\% credible regions are also similar in magnitude to the difference between the median value of \pcc\, estimated from direct simulation and the upper bound on \pcc\, from Equation \ref{eqn:bound}. This is promising for high-dimensional applications which would benefit from the decreased computational cost of the bound.
The improvement in the majority of cases by large factors suggests that this method is more sensitive and hence can result in the identification of more repeaters.

\subsection{Limitations} As discussed in Section \ref{sec:intensity}, a complicating factor is the acquisition of a proper null data set consisting of unique FRB sources detected by CHIME/FRB. The repeat bursts cannot be unambiguously removed from the sample without the results of the test for which we need the data set. However, the inclusion of repeat bursts artificially increases the intensity of the NHPP in that area and overall, making the test less likely to identify a repeater, considered conservative in this application. \rev{Another limitation of the bound specifically, as discussed in Section \ref{sec:inference}, is that for most astrophysical experiments (including CHIME/FRB) the requirement that the errors be i.i.d. is violated, although it is a reasonable assumption over the small regions tested in our \pcc\, calculation. } 


\section{Conclusions}
While much thought has gone into the treatment of nonhomogeneous Poisson processes (NHPP) observed with noise, second order characteristics of these processes had previously not been studied, since the noisy process is not typically the process of interest. In this paper, we use noisy data in a hierarchical Bayesian model to fit hyperparameters of a physically motivated nonhomogeneous Poisson process intensity. In simulation studies, we find that this methodology leads to reliable estimates of the true values of these hyperparameters. 
We then use the posterior distribution to make an inference on the noisy non-homogeneous Poisson process, the probability of detecting $k$ events within some radius $r$ and center point in the domain. 

When applied to our motivating application, the new methodology for computing probability of coincidence increases the significance of candidate repeaters previously reported in \cite{2023ApJ...947...83C} 86\% of the time, with a median improvement factor on the order of 3000. 

Our methodology has been adapted by the CHIME/FRB collaboration and will be used to define the sample for the upcoming fourth catalog of repeaters from CHIME/FRB. With an additional two years worth of data (compared to \citealt{2023ApJ...947...83C}) and the increased sensitivity offered by this methodology, the catalog is expected to roughly double the number of known repeaters. 

\rev{Additional evidence for association between the bursts could be used in tandem with \pcc. This additional evidence could be based on the likelihood of e.g., the implied rate or clustering of a source, the DM change with time, or by quantifying how similar the observed properties of the bursts are to those associated with repeaters, i.e., narrowbanded, wide, downward drifting bursts. There are many methodologies that attempt to make predictions on whether an FRB source will repeat given its phenomenological properties \citep[e.g.,][]{2022MNRAS.509.1227C, 2023MNRAS.518.1629L, 2025ApJ...982...46H}. Currently, the temporal statistics, luminosity functions, and $\Delta$DM distributions of the population of repeaters are not well known, and so their inclusion in metrics used to decide whether CHIME/FRB will announce a new repeater candidate is likely unjustified given the potential to inject bias in downstream analyses trying to answer these questions.  }

More generally, in the field of FRBs, our methodology has consequences not only for associations between FRBs (repeaters), but also between FRBs and other events (such as gravitational wave sources). 
Unambiguous classifications in these cases are vital, as increasing the repeater sample size and detecting outlier repeaters (which, for example, could be the most promising for follow-up or otherwise particularly interesting astrophysically) can give us insight into the progenitors of FRBs and the relationship between observational subclasses of FRBs. Similarly, associations between FRBs and other poorly localized transients, like gravitational wave events, which have known progenitors, could result in a field-changing discovery. \rev{Similarly, the method could be applied to sparse-event-counting astrophysical applications like neutrino, gamma-ray or X-ray observations.}

While this methodology is geared towards astrophysical observations with significant localization uncertainty, the common application of NHPPs to fields like ecology and epidemiology suggests that the method may be beneficial for research in other disciplines, for example, hotspot detection.


%
%

\begin{acks}[Acknowledgments]
We acknowledge that CHIME is located on the traditional, ancestral, and unceded territory of the Syilx/Okanagan people. We are grateful to the staff of the Dominion Radio Astrophysical Observatory, which is operated by the National Research Council of Canada.  A.M.C., R.C.J. and A.B.P. are additionally affiliated with the Trottier Space Institute. A.P. is also affiliated with the Dunlap Institute for Astronomy \& Astrophysics at the University of Toronto. 
G.M.E. is jointly appointed to the David A. Dunlap Department of Astronomy and Astrophysics at the University of Toronto. K.W.M. is also affiliated with MIT's Department of Physics.  J.X.P. is also affiliated with the Kavli Institute for the Physics and Mathematics of the Universe and National Astronomical Observatory of Japan. 
\end{acks}
\begin{funding}
This work was funded by a Canadian Statistical Sciences Institute (CANSSI) Collaborative Research Team Grant to author G.M.E.. CANSSI is supported under the Discovery Institutes Support program of the Natural Sciences and Engineering Research Council of Canada (NSERC). 
A.M.C. \rev{is a Banting Postdoctoral Research Fellow, but was funded by an NSERC Doctoral Postgraduate Scholarship while performing the majority of the analysis for this paper as a Ph.D. candidate in Astronomy at the University of Toronto.} 
D.L. acknowledges support from the Data Sciences Institute Doctoral Fellowship at the University of Toronto, and from CANSSI Ontario Multi-disciplinary Doctoral Trainee Program.
D.C.S. acknowledges the support of NSERC, RGPIN-2021-03985.
The Dunlap Institute is funded through an endowment established by the David Dunlap family and the University of Toronto. B.M.G. acknowledges the support of NSERC through grant RGPIN-2022-03163, and of the Canada Research Chairs program.
Z.P. is supported by an NWO Veni fellowship (VI.Veni.222.295).
A.P. is funded by the NSERC Canada Graduate Scholarships -- Doctoral program.
A.B.P. is a Banting, McGill Space Institute~(MSI), and FRQNT postdoctoral fellow.
J.X.P. acknowledges support from NSF grants AST-1911140, AST-1910471, and AST-2206490 as a member of the Fast and Fortunate for FRB Follow-up team.
%
CHIME is funded by a grant from the Canada Foundation for Innovation (CFI) 2012 Leading Edge Fund (Project 31170) and by contributions from the provinces of British Columbia, Qu\'{e}bec and Ontario. The CHIME/FRB Project is funded by a grant from the CFI 2015 Innovation Fund (Project 33213) and by contributions from the provinces of British Columbia and Qu\'{e}bec, and by the Dunlap Institute for Astronomy and Astrophysics at the University of Toronto. Additional support was provided by the Canadian Institute for Advanced Research (CIFAR), McGill University and the Trottier Space Institute thanks to the Trottier Family Foundation, and the University of British Columbia.
\end{funding}

\begin{supplement}
\stitle{Model evaluation and derivation of bound in Equation 7}
\sdescription{For the purposes of evaluating the goodness of fit of our  intensity model, we present and discuss raw cumulative residuals and 2D Pearson residuals. We also perform a simulation study to give context for the size of the cumulative residuals in either spatial dimension. Additionally, we present the derivation for the bound in Equation \ref{eqn:bound}.}
\end{supplement}
\begin{supplement}
\stitle{Code for MCMC and k-contact distance probability estimation}
\sdescription{We include all code required to reproduce the computed quantities in the paper.}
\end{supplement}


\bibliographystyle{imsart-nameyear} 
\bibliography{stats_bib}       

\begin{thebibliography}{53}

\bibitem[\protect\citeauthoryear{{Aggarwal} et~al.}{2021}]{2021ApJ...911...95A}
\begin{barticle}[author]
\bauthor{\bsnm{{Aggarwal}},~\bfnm{Kshitij}\binits{K.}},
  \bauthor{\bsnm{{Budav{\'a}ri}},~\bfnm{Tam{\'a}s}\binits{T.}},
  \bauthor{\bsnm{{Deller}},~\bfnm{Adam~T.}\binits{A.~T.}} \betal{et~al.}
(\byear{2021}).
\btitle{{Probabilistic association of transients to their hosts (PATH)}}.
\bjournal{\apj}
\bvolume{911}
\bpages{95}.
\bdoi{10.3847/1538-4357/abe8d2}
\end{barticle}
\endbibitem

\bibitem[\protect\citeauthoryear{Baddeley
  et~al.}{2005}]{https://doi.org/10.1111/j.1467-9868.2005.00519.x}
\begin{barticle}[author]
\bauthor{\bsnm{Baddeley},~\bfnm{A.}\binits{A.}},
  \bauthor{\bsnm{Turner},~\bfnm{R.}\binits{R.}},
  \bauthor{\bsnm{Møller},~\bfnm{J.}\binits{J.}} \betal{et~al.}
(\byear{2005}).
\btitle{Residual analysis for spatial point processes (with discussion)}.
\bjournal{Journal of the Royal Statistical Society: Series B (Statistical
  Methodology)}
\bvolume{67}
\bpages{617-666}.
\bdoi{https://doi.org/10.1111/j.1467-9868.2005.00519.x}
\end{barticle}
\endbibitem

\bibitem[\protect\citeauthoryear{{Baptista} et~al.}{2024}]{2023arXiv230507022B}
\begin{barticle}[author]
\bauthor{\bsnm{{Baptista}},~\bfnm{Jay}\binits{J.}},
  \bauthor{\bsnm{{Prochaska}},~\bfnm{J.~Xavier}\binits{J.~X.}},
  \bauthor{\bsnm{{Mannings}},~\bfnm{Alexandra~G.}\binits{A.~G.}} \betal{et~al.}
(\byear{2024}).
\btitle{{Measuring the variance of the Macquart relation in
  redshift{\textendash}extragalactic dispersion measure modeling}}.
\bjournal{\apj}
\bvolume{965}
\bpages{57}.
\bdoi{10.3847/1538-4357/ad2705}
\end{barticle}
\endbibitem

\bibitem[\protect\citeauthoryear{Bar-Hen
  et~al.}{2013}]{Bar-Hen_Chadoeuf_Dessard_Monestiez_2013}
\begin{barticle}[author]
\bauthor{\bsnm{Bar-Hen},~\bfnm{A.}\binits{A.}},
  \bauthor{\bsnm{Chadœuf},~\bfnm{J.}\binits{J.}},
  \bauthor{\bsnm{Dessard},~\bfnm{H.}\binits{H.}} \AND
  \bauthor{\bsnm{Monestiez},~\bfnm{P.}\binits{P.}}
(\byear{2013}).
\btitle{Estimating second order characteristics of point processes with known
  independent noise}.
\bjournal{Statistics and Computing}
\bvolume{23}
\bpages{297–309}.
\bdoi{10.1007/s11222-011-9311-7}
\end{barticle}
\endbibitem

\bibitem[\protect\citeauthoryear{{Beckwith} et~al.}{2006}]{2006AJ....132.1729B}
\begin{barticle}[author]
\bauthor{\bsnm{{Beckwith}},~\bfnm{Steven V.~W.}\binits{S.~V.~W.}},
  \bauthor{\bsnm{{Stiavelli}},~\bfnm{Massimo}\binits{M.}},
  \bauthor{\bsnm{{Koekemoer}},~\bfnm{Anton~M.}\binits{A.~M.}} \betal{et~al.}
(\byear{2006}).
\btitle{{The Hubble ultra deep field}}.
\bjournal{\aj}
\bvolume{132}
\bpages{1729-1755}.
\bdoi{10.1086/507302}
\end{barticle}
\endbibitem

\bibitem[\protect\citeauthoryear{Berman}{1977}]{kthnearest}
\begin{barticle}[author]
\bauthor{\bsnm{Berman},~\bfnm{Mark}\binits{M.}}
(\byear{1977}).
\btitle{Distance distributions associated with poisson processes of geometric
  figures}.
\bjournal{Journal of Applied Probability}
\bvolume{14}.
\bdoi{10.2307/3213273}
\end{barticle}
\endbibitem

\bibitem[\protect\citeauthoryear{Chakraborty and
  Gelfand}{2010}]{10.1214/10-ba504}
\begin{barticle}[author]
\bauthor{\bsnm{Chakraborty},~\bfnm{Avishek}\binits{A.}} \AND
  \bauthor{\bsnm{Gelfand},~\bfnm{Alan~E.}\binits{A.~E.}}
(\byear{2010}).
\btitle{{Analyzing spatial point patterns subject to measurement error}}.
\bjournal{Bayesian Analysis}
\bvolume{5}
\bpages{97--122}.
\bdoi{10.1214/10-ba504}
\end{barticle}
\endbibitem

\bibitem[\protect\citeauthoryear{{Chawla} et~al.}{2017}]{2017ApJ...844..140C}
\begin{barticle}[author]
\bauthor{\bsnm{{Chawla}},~\bfnm{P.}\binits{P.}},
  \bauthor{\bsnm{{Kaspi}},~\bfnm{V.~M.}\binits{V.~M.}},
  \bauthor{\bsnm{{Josephy}},~\bfnm{A.}\binits{A.}} \betal{et~al.}
(\byear{2017}).
\btitle{{A search for fast radio bursts with the GBNCC pulsar survey}}.
\bjournal{\apj}
\bvolume{844}
\bpages{140}.
\bdoi{10.3847/1538-4357/aa7d57}
\end{barticle}
\endbibitem

\bibitem[\protect\citeauthoryear{{Chen} et~al.}{2022}]{2022MNRAS.509.1227C}
\begin{barticle}[author]
\bauthor{\bsnm{{Chen}},~\bfnm{Bo~Han}\binits{B.~H.}},
  \bauthor{\bsnm{{Hashimoto}},~\bfnm{Tetsuya}\binits{T.}},
  \bauthor{\bsnm{{Goto}},~\bfnm{Tomotsugu}\binits{T.}} \betal{et~al.}
(\byear{2022}).
\btitle{{Uncloaking hidden repeating fast radio bursts with unsupervised
  machine learning}}.
\bjournal{\mnras}
\bvolume{509}
\bpages{1227-1236}.
\bdoi{10.1093/mnras/stab2994}
\end{barticle}
\endbibitem

\bibitem[\protect\citeauthoryear{{CHIME/FRB Collaboration}
  et~al.}{2018}]{overview}
\begin{barticle}[author]
\bauthor{\bsnm{{CHIME/FRB Collaboration}}} \betal{et~al.}
(\byear{2018}).
\btitle{{The CHIME fast radio burst Project: System Overview}}.
\bjournal{\apj}
\bvolume{863}
\bpages{48}.
\bdoi{10.3847/1538-4357/aad188}
\end{barticle}
\endbibitem

\bibitem[\protect\citeauthoryear{{CHIME/FRB Collaboration}
  et~al.}{2019a}]{2019Natur.566..235C}
\begin{barticle}[author]
\bauthor{\bsnm{{CHIME/FRB Collaboration}}} \betal{et~al.}
(\byear{2019}a).
\btitle{{A second source of repeating fast radio bursts}}.
\bjournal{\nat}
\bvolume{566}
\bpages{235-238}.
\bdoi{10.1038/s41586-018-0864-x}
\end{barticle}
\endbibitem

\bibitem[\protect\citeauthoryear{{CHIME/FRB Collaboration}
  et~al.}{2019b}]{2019ApJ...885L..24C}
\begin{barticle}[author]
\bauthor{\bsnm{{CHIME/FRB Collaboration}}} \betal{et~al.}
(\byear{2019}b).
\btitle{{CHIME/FRB Discovery of eight new repeating fast radio burst sources}}.
\bjournal{\apjl}
\bvolume{885}
\bpages{L24}.
\bdoi{10.3847/2041-8213/ab4a80}
\end{barticle}
\endbibitem

\bibitem[\protect\citeauthoryear{{CHIME/FRB Collaboration}
  et~al.}{2021}]{2021ApJS..257...59C}
\begin{barticle}[author]
\bauthor{\bsnm{{CHIME/FRB Collaboration}}} \betal{et~al.}
(\byear{2021}).
\btitle{{The first CHIME/FRB fast radio burst catalog}}.
\bjournal{\apjs}
\bvolume{257}
\bpages{59}.
\bdoi{10.3847/1538-4365/ac33ab}
\end{barticle}
\endbibitem

\bibitem[\protect\citeauthoryear{{CHIME Collaboration}
  et~al.}{2022}]{2022ApJS..261...29C}
\begin{barticle}[author]
\bauthor{\bsnm{{CHIME Collaboration}}} \betal{et~al.}
(\byear{2022}).
\btitle{{An overview of CHIME, the Canadian Hydrogen Intensity Mapping
  Experiment}}.
\bjournal{\apjs}
\bvolume{261}
\bpages{29}.
\bdoi{10.3847/1538-4365/ac6fd9}
\end{barticle}
\endbibitem

\bibitem[\protect\citeauthoryear{{CHIME/FRB Collaboration}
  et~al.}{2023}]{2023ApJ...947...83C}
\begin{barticle}[author]
\bauthor{\bsnm{{CHIME/FRB Collaboration}}} \betal{et~al.}
(\byear{2023}).
\btitle{{CHIME/FRB Discovery of 25 repeating fast radio burst sources}}.
\bjournal{\apj}
\bvolume{947}
\bpages{83}.
\bdoi{10.3847/1538-4357/acc6c1}
\end{barticle}
\endbibitem

\bibitem[\protect\citeauthoryear{{Cook} et~al.}{2023}]{2023ApJ...946...58C}
\begin{barticle}[author]
\bauthor{\bsnm{{Cook}},~\bfnm{Amanda~M.}\binits{A.~M.}},
  \bauthor{\bsnm{{Bhardwaj}},~\bfnm{Mohit}\binits{M.}},
  \bauthor{\bsnm{{Gaensler}},~\bfnm{B.~M.}\binits{B.~M.}} \betal{et~al.}
(\byear{2023}).
\btitle{{An FRB sent me a DM: constraining the electron column of the Milky Way
  halo with fast radio burst dispersion measures from CHIME/FRB}}.
\bjournal{\apj}
\bvolume{946}
\bpages{58}.
\bdoi{10.3847/1538-4357/acbbd0}
\end{barticle}
\endbibitem

\bibitem[\protect\citeauthoryear{{Cook} et~al.}{2025}]{suppa}
\begin{barticle}[author]
\bauthor{\bsnm{{Cook}},~\bfnm{A.~M.}\binits{A.~M.}},
  \bauthor{\bsnm{{Li}},~\bfnm{Dayi}\binits{D.}},
  \bauthor{\bsnm{{Eadie}},~\bfnm{Gwendolyn~M.}\binits{G.~M.}} \betal{et~al.}
(\byear{2025}).
\btitle{{Supplement A to $k$-contact distance for noisy nonhomogeneous spatial
  point data with application to repeating fast radio burst sources}}.
\bjournal{AOAS}
\bvolume{XX}
\bpages{XX}.
\bdoi{10.1017/to_be_entered}
\end{barticle}
\endbibitem

\bibitem[\protect\citeauthoryear{{Cordes} and
  {Lazio}}{2002}]{2002astro.ph..7156C}
\begin{barticle}[author]
\bauthor{\bsnm{{Cordes}},~\bfnm{J.~M.}\binits{J.~M.}} \AND
  \bauthor{\bsnm{{Lazio}},~\bfnm{T.~J.~W.}\binits{T.~J.~W.}}
(\byear{2002}).
\btitle{{NE2001.I. A new model for the Galactic distribution of free electrons
  and its fluctuations}}.
\bjournal{arXiv e-prints}
\bpages{astro-ph/0207156}.
\bdoi{10.48550/arXiv.astro-ph/0207156}
\end{barticle}
\endbibitem

\bibitem[\protect\citeauthoryear{{Cucala}}{2008}]{477bb175-56f0-3455-81ae-9da35286a358}
\begin{barticle}[author]
\bauthor{\bsnm{{Cucala}},~\bfnm{Lionel.}\binits{L.}}
(\byear{2008}).
\btitle{Intensity estimation for spatial point processes observed with noise}.
\bjournal{Scandinavian Journal of Statistics}
\bvolume{35}
\bpages{322--334}.
\end{barticle}
\endbibitem

\bibitem[\protect\citeauthoryear{{Curtin} et~al.}{2024}]{2024arXiv240409242C}
\begin{barticle}[author]
\bauthor{\bsnm{{Curtin}},~\bfnm{Alice~P.}\binits{A.~P.}},
  \bauthor{\bsnm{{Sirota}},~\bfnm{Sloane}\binits{S.}},
  \bauthor{\bsnm{{Kaspi}},~\bfnm{Victoria~M.}\binits{V.~M.}},
  \bauthor{\bsnm{{Tendulkar}},~\bfnm{Shriharsh~P.}\binits{S.~P.}},
  \bauthor{\bsnm{{Bhardwaj}},~\bfnm{Mohit}\binits{M.}},
  \bauthor{\bsnm{{Cook}},~\bfnm{Amanda~M.}\binits{A.~M.}},
  \bauthor{\bsnm{{Fong}},~\bfnm{Wen-Fai}\binits{W.-F.}},
  \bauthor{\bsnm{{Gaensler}},~\bfnm{B.~M.}\binits{B.~M.}},
  \bauthor{\bsnm{{Main}},~\bfnm{Robert~A.}\binits{R.~A.}},
  \bauthor{\bsnm{{Masui}},~\bfnm{Kiyoshi~W.}\binits{K.~W.}},
  \bauthor{\bsnm{{Michilli}},~\bfnm{Daniele}\binits{D.}},
  \bauthor{\bsnm{{Pandhi}},~\bfnm{Ayush}\binits{A.}},
  \bauthor{\bsnm{{Pearlman}},~\bfnm{Aaron~B.}\binits{A.~B.}},
  \bauthor{\bsnm{{Scholz}},~\bfnm{Paul}\binits{P.}} \AND
  \bauthor{\bsnm{{Shin}},~\bfnm{Kaitlyn}\binits{K.}}
(\byear{2024}).
\btitle{{Constraining near-simultaneous radio emission from short gamma-ray
  bursts using CHIME/FRB}}.
\bjournal{\apj}
\bvolume{972}
\bpages{125}.
\bdoi{10.3847/1538-4357/ad5c65}
\end{barticle}
\endbibitem

\bibitem[\protect\citeauthoryear{{ArviZ
  Developers}}{2020}]{2020ascl.soft04012A}
\begin{bmisc}[author]
\bauthor{\bsnm{{ArviZ Developers}}}
(\byear{2020}).
\btitle{{ArviZ: Exploratory analysis of Bayesian models}}.
\bhowpublished{Astrophysics Source Code Library, record ascl:2004.012}.
\end{bmisc}
\endbibitem

\bibitem[\protect\citeauthoryear{{Dolag} et~al.}{2015}]{2015MNRAS.451.4277D}
\begin{barticle}[author]
\bauthor{\bsnm{{Dolag}},~\bfnm{K.}\binits{K.}},
  \bauthor{\bsnm{{Gaensler}},~\bfnm{B.~M.}\binits{B.~M.}},
  \bauthor{\bsnm{{Beck}},~\bfnm{A.~M.}\binits{A.~M.}} \betal{et~al.}
(\byear{2015}).
\btitle{{Constraints on the distribution and energetics of fast radio bursts
  using cosmological hydrodynamic simulations}}.
\bjournal{\mnras}
\bvolume{451}
\bpages{4277-4289}.
\bdoi{10.1093/mnras/stv1190}
\end{barticle}
\endbibitem

\bibitem[\protect\citeauthoryear{{Falcke} and
  {Rezzolla}}{2014}]{2014A&A...562A.137F}
\begin{barticle}[author]
\bauthor{\bsnm{{Falcke}},~\bfnm{Heino}\binits{H.}} \AND
  \bauthor{\bsnm{{Rezzolla}},~\bfnm{Luciano}\binits{L.}}
(\byear{2014}).
\btitle{{Fast radio bursts: the last sign of supramassive neutron stars}}.
\bjournal{\aap}
\bvolume{562}
\bpages{A137}.
\bdoi{10.1051/0004-6361/201321996}
\end{barticle}
\endbibitem

\bibitem[\protect\citeauthoryear{{Fonseca} et~al.}{2020}]{2020ApJ...891L...6F}
\begin{barticle}[author]
\bauthor{\bsnm{{Fonseca}},~\bfnm{E.}\binits{E.}},
  \bauthor{\bsnm{{Andersen}},~\bfnm{B.~C.}\binits{B.~C.}},
  \bauthor{\bsnm{{Bhardwaj}},~\bfnm{M.}\binits{M.}} \betal{et~al.}
(\byear{2020}).
\btitle{{Nine new repeating fast radio burst sources from CHIME/FRB}}.
\bjournal{\apjl}
\bvolume{891}
\bpages{L6}.
\bdoi{10.3847/2041-8213/ab7208}
\end{barticle}
\endbibitem

\bibitem[\protect\citeauthoryear{Gelman and
  Rubin}{1992}]{10.1214/ss/1177011136}
\begin{barticle}[author]
\bauthor{\bsnm{Gelman},~\bfnm{Andrew}\binits{A.}} \AND
  \bauthor{\bsnm{Rubin},~\bfnm{Donald~B.}\binits{D.~B.}}
(\byear{1992}).
\btitle{{Inference from iterative simulation using multiple sequences}}.
\bjournal{Statistical Science}
\bvolume{7}
\bpages{457 -- 472}.
\bdoi{10.1214/ss/1177011136}
\end{barticle}
\endbibitem

\bibitem[\protect\citeauthoryear{Glassner}{1990}]{glassner1990graphics}
\begin{bbook}[author]
\bauthor{\bsnm{Glassner},~\bfnm{A.~S.}\binits{A.~S.}}
(\byear{1990}).
\btitle{Graphics Gems}.
\bseries{Graphics gems - IBM}.
\bpublisher{Elsevier Science}.
\end{bbook}
\endbibitem

\bibitem[\protect\citeauthoryear{{Herrera-Martin}
  et~al.}{2025}]{2025ApJ...982...46H}
\begin{barticle}[author]
\bauthor{\bsnm{{Herrera-Martin}},~\bfnm{Antonio}\binits{A.}},
  \bauthor{\bsnm{{Craiu}},~\bfnm{Radu~V.}\binits{R.~V.}},
  \bauthor{\bsnm{{Eadie}},~\bfnm{Gwendolyn~M.}\binits{G.~M.}} \betal{et~al.}
(\byear{2025}).
\btitle{{Rare Event Classification with Weighted Logistic Regression for
  Identifying Repeating Fast Radio Bursts}}.
\bjournal{\apj}
\bvolume{982}
\bpages{46}.
\bdoi{10.3847/1538-4357/adb623}
\end{barticle}
\endbibitem

\bibitem[\protect\citeauthoryear{{James}}{2023}]{2023arXiv230617403J}
\begin{barticle}[author]
\bauthor{\bsnm{{James}},~\bfnm{C.~W.}\binits{C.~W.}}
(\byear{2023}).
\btitle{{Modelling repetition in zDM: A single population of repeating fast
  radio bursts can explain CHIME data}}.
\bjournal{\pasa}
\bvolume{40}
\bpages{e057}.
\bdoi{10.1017/pasa.2023.51}
\end{barticle}
\endbibitem

\bibitem[\protect\citeauthoryear{{James} et~al.}{2022}]{2022MNRAS.509.4775J}
\begin{barticle}[author]
\bauthor{\bsnm{{James}},~\bfnm{C.~W.}\binits{C.~W.}},
  \bauthor{\bsnm{{Prochaska}},~\bfnm{J.~X.}\binits{J.~X.}},
  \bauthor{\bsnm{{Macquart}},~\bfnm{J.~P.}\binits{J.~P.}} \betal{et~al.}
(\byear{2022}).
\btitle{{The z-DM distribution of fast radio bursts}}.
\bjournal{\mnras}
\bvolume{509}
\bpages{4775-4802}.
\bdoi{10.1093/mnras/stab3051}
\end{barticle}
\endbibitem

\bibitem[\protect\citeauthoryear{{Kirsten} et~al.}{2024}]{2023arXiv230615505K}
\begin{barticle}[author]
\bauthor{\bsnm{{Kirsten}},~\bfnm{F.}\binits{F.}},
  \bauthor{\bsnm{{Ould-Boukattine}},~\bfnm{O.~S.}\binits{O.~S.}},
  \bauthor{\bsnm{{Herrmann}},~\bfnm{W.}\binits{W.}} \betal{et~al.}
(\byear{2024}).
\btitle{{A link between repeating and non-repeating fast radio bursts through
  their energy distributions}}.
\bjournal{Nature Astronomy}
\bvolume{8}
\bpages{337-346}.
\bdoi{10.1038/s41550-023-02153-z}
\end{barticle}
\endbibitem

\bibitem[\protect\citeauthoryear{Kottas and Sansó}{2007}]{KOTTAS20073151}
\begin{barticle}[author]
\bauthor{\bsnm{Kottas},~\bfnm{Athanasios}\binits{A.}} \AND
  \bauthor{\bsnm{Sansó},~\bfnm{Bruno}\binits{B.}}
(\byear{2007}).
\btitle{Bayesian mixture modeling for spatial Poisson process intensities, with
  applications to extreme value analysis}.
\bjournal{Journal of Statistical Planning and Inference}
\bvolume{137}
\bpages{3151-3163}.
\bnote{Special Issue: Bayesian Inference for Stochastic Processes}.
\bdoi{https://doi.org/10.1016/j.jspi.2006.05.022}
\end{barticle}
\endbibitem

\bibitem[\protect\citeauthoryear{{Lanman} et~al.}{2024}]{2024arXiv240207898L}
\begin{barticle}[author]
\bauthor{\bsnm{{Lanman}},~\bfnm{Adam~E.}\binits{A.~E.}},
  \bauthor{\bsnm{{Andrew}},~\bfnm{Shion}\binits{S.}},
  \bauthor{\bsnm{{Lazda}},~\bfnm{Mattias}\binits{M.}} \betal{et~al.}
(\byear{2024}).
\btitle{{CHIME/FRB Outriggers: KKO station system and commissioning results}}.
\bjournal{arXiv e-prints}
\bpages{arXiv:2402.07898}.
\bdoi{10.48550/arXiv.2402.07898}
\end{barticle}
\endbibitem

\bibitem[\protect\citeauthoryear{{Lawrence} et~al.}{2017}]{2017AJ....154..117L}
\begin{barticle}[author]
\bauthor{\bsnm{{Lawrence}},~\bfnm{Earl}\binits{E.}}, \bauthor{\bsnm{{Vander
  Wiel}},~\bfnm{Scott}\binits{S.}},
  \bauthor{\bsnm{{Law}},~\bfnm{Casey}\binits{C.}} \betal{et~al.}
(\byear{2017}).
\btitle{{The Nonhomogeneous Poisson process for fast radio burst rates}}.
\bjournal{\aj}
\bvolume{154}
\bpages{117}.
\bdoi{10.3847/1538-3881/aa844e}
\end{barticle}
\endbibitem

\bibitem[\protect\citeauthoryear{{Li} et~al.}{2021}]{2021Natur.598..267L}
\begin{barticle}[author]
\bauthor{\bsnm{{Li}},~\bfnm{D.}\binits{D.}},
  \bauthor{\bsnm{{Wang}},~\bfnm{P.}\binits{P.}},
  \bauthor{\bsnm{{Zhu}},~\bfnm{W.~W.}\binits{W.~W.}} \betal{et~al.}
(\byear{2021}).
\btitle{{A bimodal burst energy distribution of a repeating fast radio burst
  source}}.
\bjournal{\nat}
\bvolume{598}
\bpages{267-271}.
\bdoi{10.1038/s41586-021-03878-5}
\end{barticle}
\endbibitem

\bibitem[\protect\citeauthoryear{{Lund}, {Penttinen} and
  {Rudemo}}{1999}]{lund1999}
\begin{bphdthesis}[author]
\bauthor{\bsnm{{Lund}},~\bfnm{J.}\binits{J.}},
  \bauthor{\bsnm{{Penttinen}},~\bfnm{A.}\binits{A.}} \AND
  \bauthor{\bsnm{{Rudemo}},~\bfnm{M.}\binits{M.}}
(\byear{1999}).
\btitle{{Bayesian analysis of spatial point patterns from noisy observations}},
\btype{PhD thesis},
\bpublisher{Department of Mathematics and Physics, The Royal Veterinary and
  Argicultural University, Copenhagen}.
\end{bphdthesis}
\endbibitem

\bibitem[\protect\citeauthoryear{{Luo} et~al.}{2023}]{2023MNRAS.518.1629L}
\begin{barticle}[author]
\bauthor{\bsnm{{Luo}},~\bfnm{Jia-Wei}\binits{J.-W.}},
  \bauthor{\bsnm{{Zhu-Ge}},~\bfnm{Jia-Ming}\binits{J.-M.}} \betal{et~al.}
(\byear{2023}).
\btitle{{Machine learning classification of CHIME fast radio bursts - I.
  Supervised methods}}.
\bjournal{\mnras}
\bvolume{518}
\bpages{1629-1641}.
\bdoi{10.1093/mnras/stac3206}
\end{barticle}
\endbibitem

\bibitem[\protect\citeauthoryear{{Macquart} et~al.}{2020}]{2020Natur.581..391M}
\begin{barticle}[author]
\bauthor{\bsnm{{Macquart}},~\bfnm{J.~P.}\binits{J.~P.}},
  \bauthor{\bsnm{{Prochaska}},~\bfnm{J.~X.}\binits{J.~X.}},
  \bauthor{\bsnm{{McQuinn}},~\bfnm{M.}\binits{M.}} \betal{et~al.}
(\byear{2020}).
\btitle{{A census of baryons in the Universe from localized fast radio
  bursts}}.
\bjournal{\nat}
\bvolume{581}
\bpages{391-395}.
\bdoi{10.1038/s41586-020-2300-2}
\end{barticle}
\endbibitem

\bibitem[\protect\citeauthoryear{McKay, Beckman and
  Conover}{1979}]{ef76b040-2f28-37ba-b0c4-02ed99573416}
\begin{barticle}[author]
\bauthor{\bsnm{McKay},~\bfnm{M.~D.}\binits{M.~D.}},
  \bauthor{\bsnm{Beckman},~\bfnm{R.~J.}\binits{R.~J.}} \AND
  \bauthor{\bsnm{Conover},~\bfnm{W.~J.}\binits{W.~J.}}
(\byear{1979}).
\btitle{A comparison of three methods for selecting values of input variables
  in the analysis of output from a computer code}.
\bjournal{Technometrics}
\bvolume{21}
\bpages{239--245}.
\end{barticle}
\endbibitem

\bibitem[\protect\citeauthoryear{{Moroianu} et~al.}{2023}]{2023NatAs...7..579M}
\begin{barticle}[author]
\bauthor{\bsnm{{Moroianu}},~\bfnm{Alexandra}\binits{A.}},
  \bauthor{\bsnm{{Wen}},~\bfnm{Linqing}\binits{L.}},
  \bauthor{\bsnm{{James}},~\bfnm{Clancy~W.}\binits{C.~W.}} \betal{et~al.}
(\byear{2023}).
\btitle{{An assessment of the association between a fast radio burst and binary
  neutron star merger}}.
\bjournal{Nature Astronomy}
\bvolume{7}
\bpages{579-589}.
\bdoi{10.1038/s41550-023-01917-x}
\end{barticle}
\endbibitem

\bibitem[\protect\citeauthoryear{{Niu} et~al.}{2022}]{2022Natur.606..873N}
\begin{barticle}[author]
\bauthor{\bsnm{{Niu}},~\bfnm{C.~H.}\binits{C.~H.}},
  \bauthor{\bsnm{{Aggarwal}},~\bfnm{K.}\binits{K.}},
  \bauthor{\bsnm{{Li}},~\bfnm{D.}\binits{D.}} \betal{et~al.}
(\byear{2022}).
\btitle{{A repeating fast radio burst associated with a persistent radio
  source}}.
\bjournal{\nat}
\bvolume{606}
\bpages{873-877}.
\bdoi{10.1038/s41586-022-04755-5}
\end{barticle}
\endbibitem

\bibitem[\protect\citeauthoryear{{Noutsos} et~al.}{2015}]{2015A&A...576A..62N}
\begin{barticle}[author]
\bauthor{\bsnm{{Noutsos}},~\bfnm{A.}\binits{A.}},
  \bauthor{\bsnm{{Sobey}},~\bfnm{C.}\binits{C.}},
  \bauthor{\bsnm{{Kondratiev}},~\bfnm{V.~I.}\binits{V.~I.}} \betal{et~al.}
(\byear{2015}).
\btitle{{Pulsar polarisation below 200 MHz: Average profiles and propagation
  effects}}.
\bjournal{\aap}
\bvolume{576}
\bpages{A62}.
\bdoi{10.1051/0004-6361/201425186}
\end{barticle}
\endbibitem

\bibitem[\protect\citeauthoryear{{Petroff}, {Hessels} and
  {Lorimer}}{2022}]{2022A&ARv..30....2P}
\begin{barticle}[author]
\bauthor{\bsnm{{Petroff}},~\bfnm{E.}\binits{E.}},
  \bauthor{\bsnm{{Hessels}},~\bfnm{J.~W.~T.}\binits{J.~W.~T.}} \AND
  \bauthor{\bsnm{{Lorimer}},~\bfnm{D.~R.}\binits{D.~R.}}
(\byear{2022}).
\btitle{{Fast radio bursts at the dawn of the 2020s}}.
\bjournal{\aapr}
\bvolume{30}
\bpages{2}.
\bdoi{10.1007/s00159-022-00139-w}
\end{barticle}
\endbibitem

\bibitem[\protect\citeauthoryear{{Platts} et~al.}{2019}]{2019PhR...821....1P}
\begin{barticle}[author]
\bauthor{\bsnm{{Platts}},~\bfnm{E.}\binits{E.}},
  \bauthor{\bsnm{{Weltman}},~\bfnm{A.}\binits{A.}},
  \bauthor{\bsnm{{Walters}},~\bfnm{A.}\binits{A.}} \betal{et~al.}
(\byear{2019}).
\btitle{{A living theory catalogue for fast radio bursts}}.
\bjournal{\physrep}
\bvolume{821}
\bpages{1-27}.
\bdoi{10.1016/j.physrep.2019.06.003}
\end{barticle}
\endbibitem

\bibitem[\protect\citeauthoryear{{Pleunis} et~al.}{2021}]{2021ApJ...923....1P}
\begin{barticle}[author]
\bauthor{\bsnm{{Pleunis}},~\bfnm{Ziggy}\binits{Z.}},
  \bauthor{\bsnm{{Good}},~\bfnm{Deborah~C.}\binits{D.~C.}},
  \bauthor{\bsnm{{Kaspi}},~\bfnm{Victoria~M.}\binits{V.~M.}} \betal{et~al.}
(\byear{2021}).
\btitle{{Fast radio burst morphology in the first CHIME/FRB Catalog}}.
\bjournal{\apj}
\bvolume{923}
\bpages{1}.
\bdoi{10.3847/1538-4357/ac33ac}
\end{barticle}
\endbibitem

\bibitem[\protect\citeauthoryear{{Ravi} et~al.}{2023}]{2023arXiv230101000R}
\begin{barticle}[author]
\bauthor{\bsnm{{Ravi}},~\bfnm{Vikram}\binits{V.}},
  \bauthor{\bsnm{{Catha}},~\bfnm{Morgan}\binits{M.}},
  \bauthor{\bsnm{{Chen}},~\bfnm{Ge}\binits{G.}} \betal{et~al.}
(\byear{2023}).
\btitle{{Deep Synoptic Array science: a 50 Mpc fast radio burst constrains the
  mass of the Milky Way circumgalactic medium}}.
\bjournal{arXiv e-prints}
\bpages{arXiv:2301.01000}.
\bdoi{10.48550/arXiv.2301.01000}
\end{barticle}
\endbibitem

\bibitem[\protect\citeauthoryear{{Schechter}}{1976}]{1976ApJ...203..297S}
\begin{barticle}[author]
\bauthor{\bsnm{{Schechter}},~\bfnm{P.}\binits{P.}}
(\byear{1976}).
\btitle{{An analytic expression for the luminosity function for galaxies.}}
\bjournal{\apj}
\bvolume{203}
\bpages{297-306}.
\bdoi{10.1086/154079}
\end{barticle}
\endbibitem

\bibitem[\protect\citeauthoryear{{Shannon} et~al.}{2018}]{2018Natur.562..386S}
\begin{barticle}[author]
\bauthor{\bsnm{{Shannon}},~\bfnm{R.~M.}\binits{R.~M.}},
  \bauthor{\bsnm{{Macquart}},~\bfnm{J.~P.}\binits{J.~P.}},
  \bauthor{\bsnm{{Bannister}},~\bfnm{K.~W.}\binits{K.~W.}} \betal{et~al.}
(\byear{2018}).
\btitle{{The dispersion-brightness relation for fast radio bursts from a
  wide-field survey}}.
\bjournal{\nat}
\bvolume{562}
\bpages{386-390}.
\bdoi{10.1038/s41586-018-0588-y}
\end{barticle}
\endbibitem

\bibitem[\protect\citeauthoryear{{Spitler} et~al.}{2016}]{2016Natur.531..202S}
\begin{barticle}[author]
\bauthor{\bsnm{{Spitler}},~\bfnm{L.~G.}\binits{L.~G.}},
  \bauthor{\bsnm{{Scholz}},~\bfnm{P.}\binits{P.}},
  \bauthor{\bsnm{{Hessels}},~\bfnm{J.~W.~T.}\binits{J.~W.~T.}} \betal{et~al.}
(\byear{2016}).
\btitle{{A repeating fast radio burst}}.
\bjournal{\nat}
\bvolume{531}
\bpages{202-205}.
\bdoi{10.1038/nature17168}
\end{barticle}
\endbibitem

\bibitem[\protect\citeauthoryear{{Sridhar} et~al.}{2021}]{2021ApJ...917...13S}
\begin{barticle}[author]
\bauthor{\bsnm{{Sridhar}},~\bfnm{Navin}\binits{N.}},
  \bauthor{\bsnm{{Metzger}},~\bfnm{Brian~D.}\binits{B.~D.}},
  \bauthor{\bsnm{{Beniamini}},~\bfnm{Paz}\binits{P.}} \betal{et~al.}
(\byear{2021}).
\btitle{{Periodic fast radio bursts from luminous X-ray binaries}}.
\bjournal{\apj}
\bvolume{917}
\bpages{13}.
\bdoi{10.3847/1538-4357/ac0140}
\end{barticle}
\endbibitem

\bibitem[\protect\citeauthoryear{{Stenning} and {van
  Dyk}}{2018}]{2018sabm.book...29S}
\begin{bincollection}[author]
\bauthor{\bsnm{{Stenning}},~\bfnm{David~C.}\binits{D.~C.}} \AND
  \bauthor{\bsnm{{van Dyk}},~\bfnm{David~A.}\binits{D.~A.}}
(\byear{2018}).
\btitle{{Bayesian statistical nethods for astronomy. Part 2. Markov chain Monte
  Carlo}}.
In \bbooktitle{Statistics for Astrophysics: Bayesian Methodology}
(\beditor{\bfnm{Jean-Baptiste}\binits{J.-B.}~\bsnm{{Marquette}}}, ed.)
\bpages{29-58}.
\bpublisher{EDP Sciences}.
\end{bincollection}
\endbibitem

\bibitem[\protect\citeauthoryear{{Totani}}{2013}]{2013PASJ...65L..12T}
\begin{barticle}[author]
\bauthor{\bsnm{{Totani}},~\bfnm{Tomonori}\binits{T.}}
(\byear{2013}).
\btitle{{Cosmological fast radio bursts from binary neutron star mergers}}.
\bjournal{\pasj}
\bvolume{65}
\bpages{L12}.
\bdoi{10.1093/pasj/65.5.L12}
\end{barticle}
\endbibitem

\bibitem[\protect\citeauthoryear{Van~Lieshout}{2000}]{van2000markov}
\begin{bbook}[author]
\bauthor{\bsnm{Van~Lieshout},~\bfnm{M.~N.~M.}\binits{M.~N.~M.}}
(\byear{2000}).
\btitle{Markov point processes and their applications}.
\bseries{G - Reference,Information and Interdisciplinary Subjects Series}.
\bpublisher{Imperial College Press}.
\end{bbook}
\endbibitem

\bibitem[\protect\citeauthoryear{{Zhang}}{2022}]{2022arXiv221203972Z}
\begin{barticle}[author]
\bauthor{\bsnm{{Zhang}},~\bfnm{Bing}\binits{B.}}
(\byear{2022}).
\btitle{{The physics of fast radio bursts}}.
\bjournal{arXiv e-prints}
\bpages{arXiv:2212.03972}.
\bdoi{10.48550/arXiv.2212.03972}
\end{barticle}
\endbibitem

\end{thebibliography}


\end{document}